\documentclass[dandb]{article}

\usepackage[final]{neurips_2025}

\usepackage[utf8]{inputenc} 
\usepackage[T1]{fontenc}    
\usepackage{hyperref}       
\usepackage{url}            
\usepackage{booktabs}       
\usepackage{amsfonts}       
\usepackage{nicefrac}       
\usepackage{microtype}      
\usepackage{xcolor}         
\usepackage{graphicx}
\usepackage[most]{tcolorbox}
\usepackage{tabularx}
\usepackage{listings}

\newlength{\dialogboxwidth}
\newtcolorbox{systemmessage}[1][]{
  enhanced,
  title=System Prompt, 
  colback=black!3!white,    
  colframe=black!60!black,   
  fonttitle=\bfseries,
  coltitle=black,
  attach boxed title to top left={xshift=5mm, yshift*=-\tcboxedtitleheight/2},
  boxed title style={
    colback=black!20!white,
  },
  breakable,
  pad after break=2mm,
  before skip=2mm,
  after skip=2mm,
  width=\textwidth,
  #1
}

\newtcolorbox{agentmessage}[1][]{
  enhanced,
  title=Agent Output,
  colback=blue!5!white,
  colframe=blue!60!black,
  fonttitle=\bfseries,
  coltitle=black,
  attach boxed title to top left={xshift=5mm, yshift*=-\tcboxedtitleheight/2},
  boxed title style={
    colback=blue!40!white,
  },
  breakable,
  pad after break=2mm,
  before skip=2mm,
  after skip=2mm,
  width=\dialogboxwidth,
  #1
}

\newtcolorbox{envmessage}[1][]{
  enhanced,
  title=Environment Observation,
  colback=green!5!white,
  colframe=green!50!black,
  fonttitle=\bfseries,
  coltitle=black,
  attach boxed title to top left={xshift=5mm, yshift*=-\tcboxedtitleheight/2},
  boxed title style={
    colback=green!30!white,
  },
  breakable,
  pad after break=2mm,
  before skip=2mm,
  after skip=2mm,
  width=\dialogboxwidth, 
  #1
}

\title{SWE-rebench: An Automated Pipeline for Task Collection and Decontaminated Evaluation of Software Engineering Agents}

\author{%
  Ibragim Badertdinov\thanks{Equal contribution. Correspondence to \texttt{ibragim-bad@nebius.com}} \\
  Nebius \\
  \And
  Alexander Golubev\footnotemark[1] \\
  Nebius \\
  \AND
  Maksim Nekrashevich \\
  Nebius \\
  \And
  Anton Shevtsov \\
  Nebius \\
  \And
  Simon Karasik \\
  Nebius \\
  \And
  Andrei Andriushchenko \\
  Nebius \\
  \And
  Maria Trofimova \\
  Nebius \\
  \And
  Daria Litvintseva \\ 
  Nebius \\
  \And
  Boris Yangel \\
  Nebius \\
}

\begin{document}

\maketitle

\begin{abstract}
LLM-based agents have shown promising capabilities in a growing range of software engineering (SWE) tasks. However, advancing this field faces two critical challenges. First, high-quality training data is scarce, especially data that reflects real-world SWE scenarios, where agents must interact with development environments, execute code and adapt behavior based on the outcomes of their actions. Existing datasets are either limited to one-shot code generation or comprise small, manually curated collections of interactive tasks, lacking both scale and diversity. Second, the lack of fresh interactive SWE tasks affects evaluation of rapidly improving models, as static benchmarks quickly become outdated due to contamination issues. To address these limitations, we introduce a novel, automated, and scalable pipeline to continuously extract real-world interactive SWE tasks from diverse GitHub repositories. Using this pipeline, we construct SWE-rebench, a public dataset comprising over 21,000 interactive Python-based SWE tasks, suitable for reinforcement learning of SWE agents at scale. Additionally, we use continuous supply of fresh tasks collected using SWE-rebench methodology to build a contamination-free benchmark for agentic software engineering. We compare results of various LLMs on this benchmark to results on SWE-bench Verified and show that performance of some language models might be inflated due to contamination issues.
\end{abstract}

\section{Introduction}

Large Language Models (LLMs) have demonstrated impressive capabilities in SWE tasks, including code generation, debugging, and automated development workflows. Building on these capabilities, researchers have begun creating LLM-driven agents that interact with real codebases and development environments, performing actions and receiving feedback \citep{jin2025llmsllmbasedagentssoftware}. While frontier proprietary models drive the performance of the most competitive agents (e.g., OpenHands \citep{openhands}, Moatless Tools \citep{antoniades2024swesearchenhancingsoftwareagents},
Agentless \citep{xia2024agentlessdemystifyingllmbasedsoftware}) on key benchmarks like SWE-bench \citep{jimenez2024swebench}, there exists a significant opportunity to enhance open-source models \citep{wang2025-introducing-openhands-lm, yang2025swesmithscalingdatasoftware, wang2025swedev, aggarwal2025darsdynamicactionresampling, ma2025thinkinglongerlargerenhancing, wei2025swerladvancingllmreasoning, golubev2024search}. Progress in this direction, particularly toward complex agentic behaviors, may be accelerated with access to large-scale, high-quality training data that mirrors the interactivity inherent in real-world software development. Existing powerful open-source models like DeepSeek-V3 \citep{deepseekai2024deepseekv3technicalreport}, LLaMa 4 \citep{meta2025llama4} and Qwen3 \citep{qwen3} could potentially be fine-tuned to achieve comparable performance in specific SWE domains, but this hinges on the availability of suitable interactive task data.

Current approaches to training LLMs for programming often rely on code data from open-source repositories \citep{lozhkov2024starcoder2stackv2} or synthetic instruction datasets \citep{wei2024selfcodealignselfalignmentcodegeneration} that are used for instruction tuning. However, training robust software engineering agents for real-world scenarios necessitates datasets that extend beyond simple code generation. To truly enable learning through methods like Reinforcement Learning (RL), which thrives on trial-and-error, agents require interactive tasks coupled with automatic verification mechanisms. Such data must allow agents to perform diverse actions, observe environment responses after each step, and receive eventual verification outcomes that determine task success. Unlike domains such as mathematics \citep{shao2024deepseekmathpushinglimitsmathematical} or web navigation \citep{pan2024autonomousevaluationrefinementdigital}, software engineering has historically lacked such large-scale interactive datasets due to the complexities of configuring diverse, executable environments at scale. While recent efforts like SWE-Gym \citep{pan2024trainingsoftwareengineeringagents} and SWE-PolyBench \citep{rashid2025swepolybenchmultilanguagebenchmarkrepository} represent promising steps, their manual curation processes and reliance on a limited number of repositories constrain their scope, diversity, and scalability.

Furthermore, the evaluation of rapidly advancing LLM-based agents also faces significant challenges. Static benchmarks, while initially valuable, can become compromised by data contamination as newer models become exposed to test instances during their extensive pre/post-training. Moreover, the lack of standardized evaluation protocols, variability in agent scaffolds and inconsistent reporting practices make direct comparisons between models difficult and can obscure their true capabilities.

To address these challenges in both training data availability and evaluation reliability, we present a scalable, fully automated pipeline for continuous collection of software engineering tasks from real-world GitHub repositories. Building upon our prior work such as SWE-bench Extra \citep{badertdinov2024scaling}, which has been well-received by the community and is already used to train open-source software engineering agents \citep{wang2025swedev}, our approach eliminates manual intervention and significantly expands task diversity and scale. To the best of our knowledge, this is the first system enabling fully automated, scalable collection of executable tasks from a wide set of real-world repositories, specifically designed to support interactive agent training and robust benchmarking.

Our main contributions are as follows:

\begin{itemize}
    \item A scalable and fully automated pipeline for mining real-world software engineering tasks from GitHub, covering environment configuration, build setup, and test validation.
    \item SWE-rebench\footnote{Available on Hugging Face Datasets: \url{https://huggingface.co/datasets/nebius/SWE-rebench}}, a public dataset of more than 21,000 interactive Python-based SWE tasks, designed to train and benchmark agents in diverse executable environments, particularly suitable for reinforcement learning-based approaches.
    \item A public SWE-rebench leaderboard\footnote{Accessible at: \url{https://swe-rebench.com/leaderboard}} that offers continuously updated, decontaminated, and standardized evaluations for LLM-based agents, promoting transparency and fair comparisons across both open- and closed-source models. As the leaderboard is continuously updated, the current set of evaluated models can be found on the leaderboard website.
\end{itemize}

By focusing on scale and automation, SWE-rebench aims to fill a critical gap in the LLM agent ecosystem. We believe it will serve as a foundational resource for accelerating open-source research and improving the reliability and performance of LLM-based software engineering agents.

\section{An automated pipeline for collecting software engineering tasks}
\label{sec:automated_pipeline}

In this section we describe our automated pipeline for mining verifiable software engineering tasks at scale, that we used to build SWE-rebench, a dataset of 21,336 verifiable SWE tasks from 3468 distinct GitHub repositiories. Our pipeline comprises four stages: \textit{preliminary task collection}, \textit{automated installation instruction configuration}, \textit{execution-based installation verification}, and \textit{quality assessment}, which are fully described in this section. While our methodology incorporates several techniques from SWE-bench, it also introduces innovations to enhance automation and scalability. We detail the distinctions and novel aspects of our approach compared to the original SWE-bench methodology in Appendix~\ref{sec:list_of_refinements}. The computationally intensive nature of our pipeline is managed through a distributed storage and computing platform TractoAI \citep{tractoai2025}, which provides capabilities for efficient parallel processing and data management, helping us optimize throughput of each stage to enable rapid reprocessing whenever we change the pipeline.

\subsection{Preliminary task collection}
\label{sec:prelim_task_collection}

In the first stage, we download raw input data from multiple origins, merge them, and perform preliminary filtering. The primary sources for our data are GitHub Archive \citep{grigorik2011gharchive} and GitHub.

\begin{itemize}
    \item \textbf{GitHub Archive.} The GitHub Archive is a major source of public events on GitHub. Each day, it publishes a JSON archive listing all GitHub events from that day. We use this archive to collect detailed data about issues: issue description, discussion, linked pull requests, and metadata such as creation date and labels. We also extract information about pull requests, including their merge status, last commit, and discussions.
    \item \textbf{GitHub.} We clone relevant GitHub repositories with their full commit histories to our local storage. A local copy enables efficient access to repository data and helps avoid GitHub API rate limits. We use preserved commit history to identify changes associated with pull requests and perform version analysis for automated dependency setup in later stages.
\end{itemize}

To initiate the dataset building process, we download approximately 450,000 pull requests linked to issues created before May 1, 2025. These originate from over 30,000 repositories that feature permissive licenses granting broad usage rights (see Appendix~\ref{appendix:licenses} for the list of included license types) and where Python constitutes over 75\% of the codebase lines of code.

We then link issues with pull requests that mention resolving them in their title or description, applying filters to select instances where:
\begin{itemize}
    \item The issue is from a Python repository with a permissive license.
    \item The issue is marked as resolved.
    \item The PR is merged into the main branch.
    \item The PR is not linked to multiple issues.
    \item The issue description is longer than 10 characters.
    \item The PR must introduces or modifies tests and includes code changes beyond test files.
    \item Changes affect 1 to 15 files.
\end{itemize}

This filtering aims to eliminate unsuitable candidates, particularly those lacking tests. We require pull requests that introduce or modify tests, as these are crucial for automatically evaluating whether a proposed code patch resolves the described issue. For each selected pull request, the overall patch is divided into two components: a solution patch, containing changes to non-test files intended to address the issue, and a test patch, comprising only changes to test files. After applying all filtering criteria, approximately 153,400 potential task instances remain.

\begin{figure}
    \centering
    \includegraphics[width=0.9\linewidth]{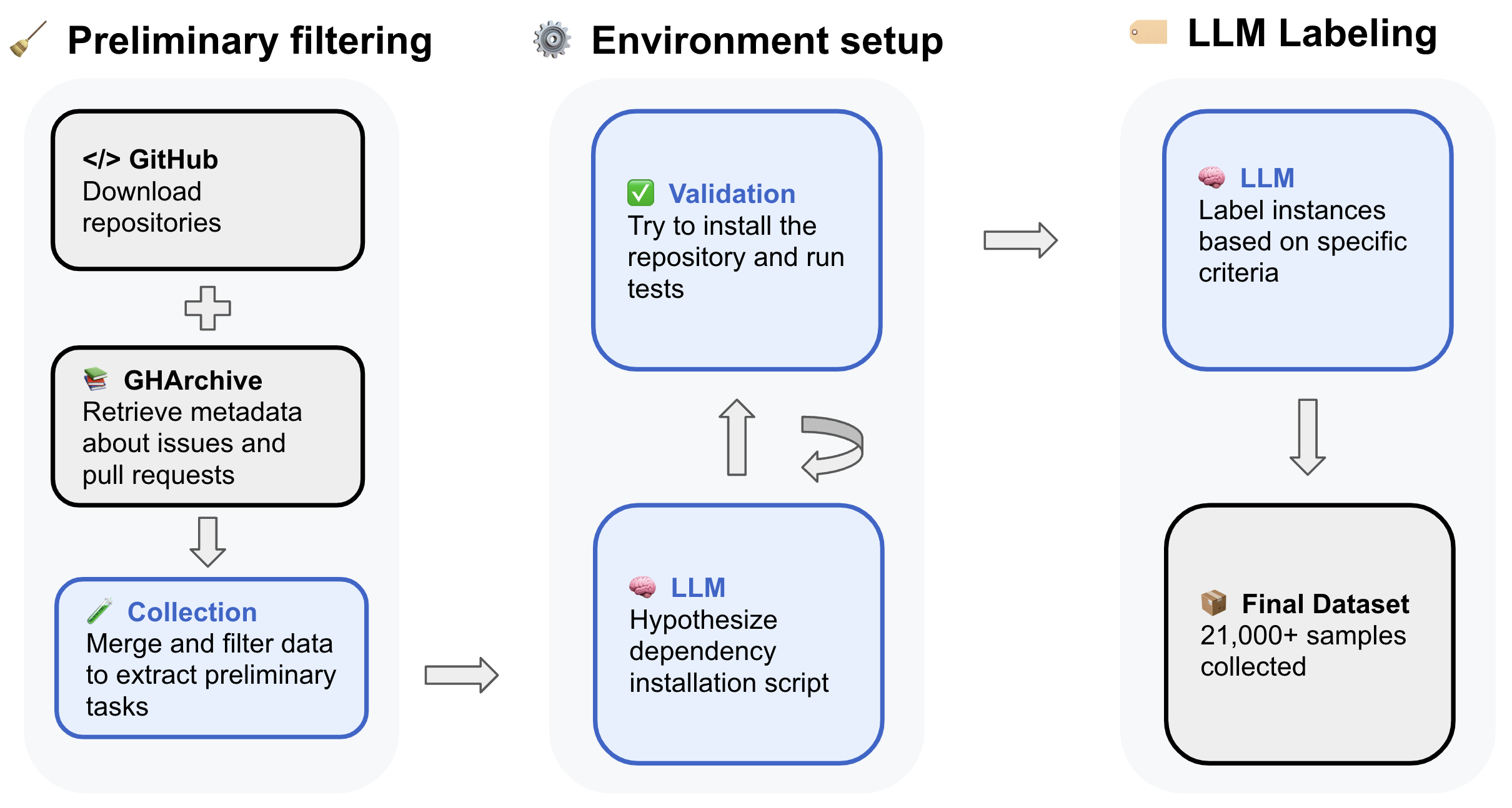}
    \caption{Overview of the automated pipeline for collecting software engineering data.}
    \label{fig:data_pipeline_overview}
\end{figure}

\subsection{Automated installation instructions configuration}
\label{sec:automated_install_conf}

Datasets like SWE-bench \citep{jimenez2024swebench} or SWE-Gym \citep{pan2024trainingsoftwareengineeringagents} rely on manual curation to configure executable environments for each repository. This approach inherently limits scalability, often confining such datasets to a small selection of well-known repositories. Key steps typically include project versioning (mapping multiple task instances to a single valid environment) and defining setup instructions (to install dependencies and run tests). Manually conducting these steps on a large-scale, diverse task collection is infeasible; therefore, we employ a fully automated approach.

After preliminary filtering described in Section \ref{sec:prelim_task_collection}, remaining issues are treated as task instances. We group these task instances by project versions inferred from \verb|git tag| outputs, normalizing versions to \verb|major.minor| format (e.g., \verb|1.2.3| is normalized to \verb|1.2|). For each version group, we select the \verb|base_commit| of the pull request linked to the task instance with the most recent \verb|base_commit| date. We prioritize this most recent \verb|base_commit| because developers typically maintain dependency compatibility within minor versions, and later commits often include important environment fixes. This approach generally provides a stable dependency set, often sufficient for executing test patches from all tasks in that group within a shared environment. The \verb|git tag| command provides a version for approximately 95\% of task instances. We assign a unique version to the rest of the tasks, so that each one of them uses its own environment.
We employ an agentless approach, inspired by \citep{xia2024agentlessdemystifyingllmbasedsoftware}, to generate candidate environment setup instructions. This process involves several LLM-driven steps:

\begin{itemize}
    \item \textbf{Identifying relevant files:} An LLM scans repository files (e.g., \verb|README.md|, \verb|Dockerfile|, \verb|setup.py|) to find potential sources of installation information.
    \item \textbf{Extracting installation recipe:} The LLM processes the concatenated content of files identified in the previous stage to produce a structured JSON object detailing the installation recipe. Files are provided to the LLM in the format: \verb|<filename>F.ext</filename>\n<content>CONTENT</content>|. An example of the LLM's reasoning and the resulting JSON recipe is provided in Appendix~\ref{appendix:json_recipe_example}.
\end{itemize}

We use the \texttt{Qwen2.5-72B-Instruct} model \citep{qwen2025qwen25technicalreport} (prompt in Appendix~\ref{sec:extract_install_recipe}) to generate up to three candidate JSON recipes per task. If an error occurs during the subsequent scripted installation or test execution (derived from a recipe), the LLM attempts to refine that recipe by analyzing error logs and the original instructions (see correction prompt in Appendix~\ref{sec:update_install_recipe}). This iterative refinement enables successful environment configuration for tasks with issues like missing libraries or incorrect setups, allowing their inclusion in the final dataset. Our approach successfully produces a working installation recipe for at least one task in 31\% of all repositories. We also catalog the primary installation failure reasons (e.g., unpinned dependencies or flaky tests) and provide a brief review in Appendix~\ref{appendix:failure_modes}.

We also explored dependency installation using an interactive agent that directly interacts with a Docker environment to install projects and run tests. While this interactive agent occasionally configured environments more effectively, it proved to be significantly more resource-demanding. The chosen agentless method is more computationally efficient for large-scale processing, and generating multiple candidate recipes can further improve its effectiveness, making it our primary approach. A comparative evaluation of these approaches on a curated subset of SWE-bench tasks is detailed in Appendix~\ref{appendix:install_method_eval}.

\subsection{Execution-based installation verification}
\label{sec:exec_based_verification}

To confirm task solvability and the integrity of the provided tests, we perform execution-based installation verification. This stage involves installing the environment for each task within a container and executing the tests from the pull request's test patch. We then parse the test run outputs to ensure that: (1) at least one test from the test patch fails before applying the solution patch (i.e., changes to non-test files from the original pull request), (2) all tests from the test patch that initially failed subsequently pass after the solution patch is applied, and (3) any tests from the test patch that initially passed continue to pass after the solution patch is applied. Tasks are considered valid only if they meet these conditions.

Processing numerous task instances, each potentially with multiple candidate recipes requiring installation, testing, and logging, necessitates distributed execution to manage the workload efficiently. We use TractoAI for this purpose, as it enables distributed container building and parallel execution of verification tasks in built containers across a cluster.

For installation verification, we use a default base container with pre-installed basic dependencies (e.g., \textbf{conda}, \textbf{gcc}). Our own image registry and internal PyPI/APT mirrors help cache popular dependencies, accelerating container launches and reducing reliance on external sources. During task verification we perform the following steps:

\begin{itemize}
    \item Install project dependencies in an isolated container using \textbf{buildah}. We utilize \textbf{tmpfs} for file system operations to minimize disk I/O and accelerate builds.
    \item Execute tests and parse logs to identify tests validating the solution.
    \item Build final container images upon successful verification.
    \item Record exact dependency versions (via \verb|pip freeze|, \verb|conda env export|) after a successful setup to mitigate reproducibility issues from unpinned dependencies in Python projects, ensuring consistent environment recreation in the future.
\end{itemize}

\subsection{Automated instance quality assessment}
\label{sec:automated_quality_assess}

For our collected tasks to be effectively used for reinforcement learning, they should possess certain properties; otherwise, RL agents might generate trajectories that appear as failures but are actually due to task imperfections (e.g., an underspecified issue making the task unsolvable, or flawed tests that a correct solution cannot pass), leading to incorrectly penalizing the agent. While SWE-bench Verified ensures these properties through manual verification, the scale of our collection necessitates an automated approximation of these checks. To assess these properties automatically, we fine-tune an instruction-following model using human annotations from SWE-bench Verified to predict:
\begin{itemize}
    \item \textbf{Issue Clarity:} Whether the GitHub issue description is sufficiently detailed for a developer to understand and solve the problem.
    \item \textbf{Task Complexity:} The estimated effort to resolve the issue, considering reasoning, code modification, and codebase familiarity.
    \item \textbf{Test Patch Correctness:} Whether tests in the pull request accurately verify the intended fix without over-reliance on specific implementation details.
\end{itemize}

We fine-tune Qwen 2.5-72B-Instruct using annotations from SWE-bench Verified. For each of the over 3,800 examples, the model receives the issue description, the canonical solution patch, and the test patch as input. It is then prompted to predict one of three binary quality labels: \textbf{Issue Clarity}, \textbf{Task Complexity}, or \textbf{Test Patch Correctness}. We train the model to predict each label independently; each task instance is assessed for each quality characteristic separately using a 75/25 training/validation split (total 413 validation examples).

For \textbf{Task Complexity} (where 'high-score' implies >1 hour to solve; 100 high-score vs. 313 low-score examples in validation), our fine-tuned model achieved 81\% accuracy and a weighted F1-score of 0.82. This is an improvement over the baseline Qwen-72B-Instruct, which achieved 68\% accuracy. For \textbf{Test Patch Correctness} (180 high-score vs. 233 low-score examples), the model achieved 67\% accuracy (weighted F1: 0.65). For \textbf{Issue Clarity} (84 high-score vs. 329 low-score examples), it achieved 79\% accuracy (weighted F1: 0.76). A more detailed prediction quality analysis, including precision and recall per class, can be found in Appendix~\ref{sec:auto_label_metrics}, Table~\ref{tab:alltasks_fullreport}.

The LLM-generated labels for \textbf{Issue Clarity}, \textbf{Task Complexity}, and \textbf{Test Patch Correctness} are provided as metadata with each task instance. While this automated assessment is not perfect, these labels offer users a means to filter the dataset and select task instances according to their specific criteria. For example, these labels facilitate task difficulty control more precise than heuristics like the number of modified files (used, for example, for SWE-bench Lite subset), as the number of changed files can be misleading: a multi-file change might be simple (e.g., a repeated parameter update), while a single-file change might lack clear issue descriptions or adequate tests for full validation. Thus, these LLM-based scores for difficulty and clarity of description and tests empower users to perform more nuanced task selection, helping them identify challenging yet solvable and clearly specified tasks beneficial for their specific model training or evaluation needs, and potentially aiding in mitigating benchmark saturation.

This four-stage pipeline automates the collection and processing of interactive software engineering tasks. The process yields the SWE-rebench dataset of 21,336 annotated task instances, which is publicly available on \href{https://huggingface.co/datasets/nebius/SWE-rebench}{Hugging Face Datasets}. Accompanying code for utilizing the dataset, including scripts for tasks evaluation, is available on \href{https://github.com/SWE-rebench/SWE-bench-fork}{GitHub}. An example of a task instance with its full annotation is provided in Appendix~\ref{appendix:task_example}. 

\section{SWE-rebench benchmark}
\label{sec:swe_rebench_benchmark}

In this section we discuss key limitations of existing evaluation setups for LLM-based software engineering agents and how our automated data pipeline described in Section~\ref{sec:automated_pipeline} helps to address them. We leverage this pipeline to construct \href{http://swe-rebench.com}{SWE-rebench}, a benchmark built from hundreds of real-world, executable SWE tasks.
It comprises 294 executable tasks from 169 diverse repositories, selected using filtering criteria detailed in Appendix~\ref{appendix:filtering_criteria}, and is part of the broader SWE-rebench dataset release. To ensure reliable and standardized evaluation, we maintain a private leaderboard based on this benchmark.

\subsection{Challenges in SWE agent benchmarking}

We identified the following key areas for improvement:

\begin{itemize}
    \item \textbf{Potential data contamination:} SWE-bench, the de facto evaluation standard for SWE agents, has been public since late 2023. Models released afterward may have been exposed to its data during training, risking inflated scores and confounding generalization with memorization.
    \item \textbf{Incomparable results due to scaffolding variability:} Current evaluation practices allow for a wide range of setups. Performance on SWE-bench is often heavily influenced by highly engineered prompts, complex multi-agent frameworks, retry mechanisms, best-of-N sampling strategies and validation loops. While these techniques demonstrate the potential of systems built around LLMs, they make it difficult to isolate and compare raw capabilities of different LLMs. Furthermore, the scaffoldings are often developed and tuned on subsets from SWE-bench, inadvertently leading to a potential for implicit overfitting to the benchmark’s specific characteristics.
    \item \textbf{Lack of standardized and verifiable evaluation:} SWE-bench results are typically performed and reported by individual teams. This decentralized approach lacks a mechanism for independent verification and can potentially lead to inconsistencies or misleading reporting practices such as reporting pass@N as pass@1 or implicitly using information derived from final tests. The reliance on closed-source frameworks for many submissions further reduces the transparency and reproducibility of the evaluation process.
    \item \textbf{High variance in agent performance across runs:} Due to the stochastic nature of agent trajectories, the outcome of a single run can vary significantly. This includes cases where a model may successfully generate correct actions or recover from mistakes in some runs, but fail to do so in others. Without averaging or reporting performance across multiple runs, the results can be unrepresentative. In particular, evaluating an agent multiple times and reporting only the best-performing run risks overstating the model’s actual capabilities and resolved rate.
\end{itemize}

\subsection{Principles of SWE-rebench}

SWE-rebench is designed to address the above challenges and support rigorous, model-centric evaluation through several core principles:

\begin{itemize}
    \item \textbf{Centralized and standardized evaluation framework:} All evaluations on SWE-rebench are conducted by our team by using a fixed scaffolding, i.e., every model is assessed by using the same minimal ReAct-style agentic framework \citep{yao2023reactsynergizingreasoningacting}, identical prompts and default generation hyperparameters as recommended by model developers. We standardize the context length to 128K tokens for all evaluations, unless a model only supports a shorter context. \newline
    This strict standardization ensures an equal environment, allowing for direct comparison of the core abilities of different models to understand and solve SWE tasks within a defined, general-purpose interaction structure. While model-specific tuning or a different scaffolding could potentially yield higher scores for a given model, our focus is on establishing a reliable baseline of model capabilities in a common setting. \newline
    It’s important to note that the interaction with the development environment is based on the model generating textual commands according to the interaction format described in the prompt. To equalize evaluations, we don’t use the function-calling functionality that some of the tested models support. For transparency, we share the exact system prompt used for all model evaluations in Appendix~\ref{appendix:system_prompt}.
    \item \textbf{Continuous dataset updates and decontamination:} SWE-rebench uses an automated pipeline from Section \ref{sec:automated_pipeline} for a continuous supply of fresh tasks. Since we precisely track the creation dates of the issues and their corresponding pull requests against model release dates, we can explicitly mark potentially contaminated evaluations that include issues created before a model’s release date. These evaluations are explicitly marked on our leaderboard, to ensure transparency around possible data leakage.
    \item \textbf{Accounting for stochasticity in agent behavior:} To capture performance variability, we run each model five times on the full benchmark. We additionally report both the standard error of the mean (SEM) and pass@5 metrics to provide a statistically grounded and more reliable assessment of each model performance.
\end{itemize}

This standardized approach allows SWE-rebench to focus on measuring two fundamental aspects of model performance:

\begin{itemize}
    \item The ability to comprehend a real-world software issue (presented as a GitHub issue), devise a plan, implement a correct code patch, and potentially validate the solution.
    \item The ability to follow instructions and operate within a structured agentic framework, which is represented by our ReAct scaffolding.
\end{itemize}

\subsection{Result analysis}

\begin{table}
  \caption{Comparison of model performance on SWE-rebench Jan 2025 and SWE-rebench (Mar–Apr 2025). All metrics are reported in percentages. Models released after 1st of March 2025 are denoted with an asterisk (*).}
  \label{tab:swe-rebench-Q1-vs-Q2}
  \centering
  \begin{tabular}{lcccccc}
    \toprule
    Model & 
    \multicolumn{3}{c}{SWE-rebench (Jan)} & 
    \multicolumn{3}{c}{SWE-rebench (Mar–Apr)} \\
    \cmidrule(r){2-4} \cmidrule(r){5-7}
    & Resolved & SEM & Pass@5 & Resolved & SEM & Pass@5 \\
    \midrule
    gpt-4.1-2025-04-14* & 31.1 & 0.79 & 44.4 & 26.7 & 1.09 & 39.0 \\
    DeepSeek-V3-0324* & 21.7 & 1.64 & 35.0 & 21.3 & 0.98 & 32.4 \\
    DeepSeek-V3-1226 & 19.1 & 0.58 & 33.3 & 21.9 & 1.44 & 31.4 \\
    Qwen3-235B-A22B no-think* & 15.2 & 1.76 & 29.9 & 16.6 & 0.93 & 25.7 \\
    Qwen3-235B-A22B think* & 13.7 & 1.81 & 29.9 & 12.2 & 1.33 & 25.7 \\
    Qwen3-32B no-think* & 13.2 & 1.17 & 23.9 & 13.7 & 1.03 & 26.7 \\
    Qwen3-32B think* & 11.8 & 0.99 & 22.2 & 11.2 & 0.56 & 21.0 \\
    Llama-4-Maverick-Instruct* & 8.5 & 0.90 & 20.5 & 12.2 & 1.69 & 27.6 \\
    Qwen2.5-72B-Instruct & 8.2 & 0.79 & 18.8 & 9.3 & 1.26 & 19.0 \\
    Llama-3.3-70B-Instruct & 8.2 & 0.44 & 15.4 & 11.2 & 0.47 & 22.9 \\
    Llama-4-Scout-Instruct* & 5.0 & 0.63 & 12.8 & 5.3 & 0.38 & 14.3 \\
    gemma-3-27b-it* & 4.3 & 0.81 & 9.4 & 4.8 & 0.30 & 10.5 \\
    Qwen2.5-Coder-32B-Instruct & 2.7 & 0.57 & 7.7 & 3.2 & 0.77 & 9.5 \\
    \bottomrule
  \end{tabular}
\end{table}

We leverage the decontaminated nature of SWE-rebench to analyze performance trends over time and identify potential signs of contamination effects in prior benchmarks. Specifically, we evaluate models on two distinct temporal subsets of tasks: those created in January 2025 and those from March–April 2025. Table \ref{tab:swe-rebench-Q1-vs-Q2} presents model performance across these time windows.

To investigate potential overfitting to the SWE-bench Verified dataset, we compare model performance on SWE-rebench tasks to the same models’ performance on SWE-bench Verified. This comparison focuses on open-source models released in 2024 or early 2025, for which the risk of data leakage from the Verified subset is higher. Table \ref{tab:swe-rebench-vs-verified} summarizes the comparative results on SWE-bench Verified and the March-April 2025 slice of SWE-rebench.

The results from this evaluation showcase several notable observations:

\begin{itemize}
    \item GPT-4.1 is the only model, which performance noticeably declined on the March–April subset compared to the January subset.
    \item LLaMa-4-Maverick exhibits a high pass@5 score relative to models with similar mean resolution rates, yet has a relatively modest resolution rate. This indicates that while the model can produce correct solutions to more complex problems, it lacks reliability across runs, demonstrating high potential but inconsistent execution.
    \item Qwen2.5-Coder-32B-Instruct underperforms expectations, especially considering its strong code generation capabilities. Analysis of its trajectories reveals problems with instruction following; the model frequently hallucinated environment responses or enters loops of formatting errors, ultimately failing without producing a meaningful solution attempt.
    \item Qwen3 models perform similarly with or without think mode enabled -- in some cases, the no-think variant even slightly surpasses the think version. This suggests the base model’s capabilities are sufficiently strong for deliberate planning to provide no measurable advantage. The nearly identical pass@5 scores further indicate that the model’s problem-solving efficiency remains consistent even without explicit reasoning mechanisms
    \item DeepSeek models demonstrate the strongest performance among open-source models across both SWE-rebench subsets and the SWE-bench Verified benchmark. Notably, both the December and March releases of DeepSeek-V3 consistently outperform other open models in resolution rate and pass@5, highlighting their robustness to changes in task distribution. Interestingly, while both versions perform similarly on SWE-rebench, their scores diverge on SWE-bench Verified, which may suggest potential contamination effects on the older benchmark.
\end{itemize}

\begin{table}
  \caption{Comparison of model performance on SWE-bench Verified and SWE-rebench (Mar–Apr 2025). All metrics are reported in percentages.}
  \label{tab:swe-rebench-vs-verified}
  \centering
  \begin{tabular}{lcccccc}
    \toprule
    Model & 
    \multicolumn{3}{c}{SWE-bench Verified} & 
    \multicolumn{3}{c}{SWE-rebench (Mar–Apr)} \\
    \cmidrule(r){2-4} \cmidrule(r){5-7}
    & Resolved & SEM & Pass@5 & Resolved & SEM & Pass@5 \\
    \midrule
    DeepSeek-V3-0324 & 39.7 & 0.35 & 57.4 & 21.3 & 0.98 & 32.4 \\
    DeepSeek-V3-1226      & 35.2 & 0.52 & 51.0 & 21.9 & 1.44 & 31.4 \\
    LLaMA-3.3-70B-Instruct & 18.1 & 0.66 & 31.6 & 11.2 & 0.47 & 22.9 \\
    LLaMA-4-Maverick-Instruct & 16.0 & 0.79 & 39.2 & 12.2 & 1.69 & 27.6 \\
    Qwen2.5-72B-Instruct   & 11.3 & 0.35 & 27.8 & 9.3 & 1.26 & 19.0 \\
    LLaMA-4-Scout-Instruct & 8.8 & 0.48 & 22.2 & 5.3 & 0.38 & 14.3 \\
    Qwen2.5-Coder-32B-Instruct & 4.9 & 0.46 & 16.0 & 3.2 & 0.77 & 9.5 \\
    \bottomrule
  \end{tabular}
\end{table}

For evaluation details and experimental setup, see Appendix~\ref{appendix:exp_setup}.

\section{Discussion and limitations}
\label{sec:discussion_and_limitations}

Our automated pipeline and the resulting SWE-rebench dataset are designed to address the lack of large-scale, real-world tasks for agent-based training, and the need for up-to-date benchmarks that remain free from data contamination. By automating the extraction and validation of executable tasks, we enable broad coverage and continual supply of fresh data. However, the emphasis on scalability introduces trade-offs, particularly a reduced ability to manually curate and verify the quality and clarity of each individual task.

Extracting consistently high-quality, verifiable SWE tasks from diverse real-world GitHub repositories (Section~\ref{sec:automated_pipeline}) is an inherently imperfect process. While our multi-stage filtering, refinements to existing methodologies and automated dependency installation are designed for robustness at scale, they rely on heuristics and LLM-driven interpretations. For instance, our LLM-based approach to generating installation instructions from repository files (Qwen2.5-72B-Instruct, Section~\ref{sec:automated_install_conf}), while far more scalable than manual methods, was validated on a limited set of 18 repositories for prompt engineering and may not capture every project's subtleties. Similarly, the automated task quality assessment (Section~\ref{sec:automated_quality_assess}), where an LLM is fine-tuned on SWE-bench Verified task labels to predict complexity and relevance, serves as a valuable scalable proxy but cannot fully replicate nuanced human judgment, thus, containing errors and decreasing quality of the datasets.

Finally, while our benchmark is intended to support transparency and standardization in evaluating SWE agents, it may also accelerate the development of increasingly autonomous AI systems in software engineering. This progress brings potential risks, such as overreliance on AI-generated code or misuse of automated agents for introducing vulnerabilities. We believe that fostering openness, decontaminated evaluations, and rigorous benchmarking practices helps mitigate these concerns and contributes to responsible advancement of the field.

We outline following main limitations of our work:

\begin{itemize}
    \item \textbf{Automated task quality assessment:} While we employ automated quality assessment, the fully automated pipeline may result in some tasks being imperfectly described or unsolvable solely from the issue. This can lead to lower absolute success rates compared to manually curated benchmarks.
    \item \textbf{Limited language diversity:} The initial release of SWE-rebench is limited exclusively to Python, restricting its immediate applicability to other language ecosystems. Although the underlying pipeline is language-agnostic, adapting it to support other languages (such as Go, Java, Rust, JavaScript/TypeScript) requires implementing language-specific components, which is planned for future work.
\end{itemize}

\section{Conclusion and future work}

We have introduced a novel, fully automated pipeline for the continuous collection of software engineering tasks from open-source repositories. This pipeline provides a scalable and reliable source of fresh, decontaminated data, particularly suitable for training and evaluating LLM-based agents. It serves as the foundation for SWE-rebench public datasets and a continuously updated benchmark designed for robust and transparent evaluation of agent performance in realistic SWE scenarios. By shifting the paradigm towards automated data collection and dynamic benchmarking, we address critical limitations of existing static benchmarks and thereby facilitate more rapid and open progress in the field of AI for software engineering. Our future work will concentrate on several key areas:

\begin{itemize}
    \item \textbf{Expanding data coverage and volume:} We aim to significantly increase the dataset volume by extending our collection methodology from tasks strictly tied to GitHub issues to a broader set of code changes represented by arbitrary pull requests.
    \item \textbf{Improving task filtering pipeline:} To improve the overall quality of extracted tasks, we aim to refine our filtering heuristics used in the pipeline.
    \item \textbf{Support for new programming languages:} Applying the same core methodology, we plan to collect datasets for projects in other popular languages (e.g., JavaScript, Java, C++), thereby expanding the linguistic and technological diversity of SWE-rebench.
    \item \textbf{Keeping SWE-rebench benchmark up-to-date:} Maintaining evaluation process on fresh tasks for the existing models, evaluating a wider range of LLMs and sharing detailed performance analyses with the community.
\end{itemize}

We believe that our automated data collection pipeline and the SWE-rebench benchmark provide a vital foundation for developing and assessing the next generation of LLM-based agents for real-world software engineering challenges.

\bibliographystyle{unsrtnat}
\bibliography{neurips_2025}

\begin{thebibliography}{43}
\providecommand{\natexlab}[1]{#1}
\providecommand{\url}[1]{\texttt{#1}}
\expandafter\ifx\csname urlstyle\endcsname\relax
  \providecommand{\doi}[1]{doi: #1}\else
  \providecommand{\doi}{doi: \begingroup \urlstyle{rm}\Url}\fi

\bibitem[Jin et~al.(2025)Jin, Huang, Cai, Yan, Li, and Chen]{jin2025llmsllmbasedagentssoftware}
Haolin Jin, Linghan Huang, Haipeng Cai, Jun Yan, Bo~Li, and Huaming Chen.
\newblock From llms to llm-based agents for software engineering: A survey of current, challenges and future, 2025.
\newblock URL \url{https://arxiv.org/abs/2408.02479}.

\bibitem[Wang et~al.(2024)Wang, Li, Song, Xu, Tang, Zhuge, Pan, Song, Li, Singh, Tran, Li, Ma, Zheng, Qian, Shao, Muennighoff, Zhang, Hui, Lin, Brennan, Peng, Ji, and Neubig]{openhands}
Xingyao Wang, Boxuan Li, Yufan Song, Frank~F. Xu, Xiangru Tang, Mingchen Zhuge, Jiayi Pan, Yueqi Song, Bowen Li, Jaskirat Singh, Hoang~H. Tran, Fuqiang Li, Ren Ma, Mingzhang Zheng, Bill Qian, Yanjun Shao, Niklas Muennighoff, Yizhe Zhang, Binyuan Hui, Junyang Lin, Robert Brennan, Hao Peng, Heng Ji, and Graham Neubig.
\newblock {OpenHands: An Open Platform for AI Software Developers as Generalist Agents}, 2024.
\newblock URL \url{https://arxiv.org/abs/2407.16741}.

\bibitem[Antoniades et~al.(2024)Antoniades, Örwall, Zhang, Xie, Goyal, and Wang]{antoniades2024swesearchenhancingsoftwareagents}
Antonis Antoniades, Albert Örwall, Kexun Zhang, Yuxi Xie, Anirudh Goyal, and William Wang.
\newblock Swe-search: Enhancing software agents with monte carlo tree search and iterative refinement, 2024.
\newblock URL \url{https://arxiv.org/abs/2410.20285}.

\bibitem[Xia et~al.(2024)Xia, Deng, Dunn, and Zhang]{xia2024agentlessdemystifyingllmbasedsoftware}
Chunqiu~Steven Xia, Yinlin Deng, Soren Dunn, and Lingming Zhang.
\newblock Agentless: Demystifying llm-based software engineering agents, 2024.
\newblock URL \url{https://arxiv.org/abs/2407.01489}.

\bibitem[Jimenez et~al.(2024)Jimenez, Yang, Wettig, Yao, Pei, Press, and Narasimhan]{jimenez2024swebench}
Carlos~E Jimenez, John Yang, Alexander Wettig, Shunyu Yao, Kexin Pei, Ofir Press, and Karthik~R Narasimhan.
\newblock {SWE}-bench: Can language models resolve real-world github issues?
\newblock In \emph{The Twelfth International Conference on Learning Representations}, 2024.
\newblock URL \url{https://openreview.net/forum?id=VTF8yNQM66}.

\bibitem[Wang(2025)]{wang2025-introducing-openhands-lm}
Xingyao Wang.
\newblock Introducing openhands lm 32b -- a strong, open coding agent model.
\newblock \emph{All Hands AI Blog}, March 2025.
\newblock URL \url{https://www.all-hands.dev/blog/introducing-openhands-lm-32b----a-strong-open-coding-agent-model}.

\bibitem[Yang et~al.(2025{\natexlab{a}})Yang, Leret, Jimenez, Wettig, Khandpur, Zhang, Hui, Press, Schmidt, and Yang]{yang2025swesmithscalingdatasoftware}
John Yang, Kilian Leret, Carlos~E. Jimenez, Alexander Wettig, Kabir Khandpur, Yanzhe Zhang, Binyuan Hui, Ofir Press, Ludwig Schmidt, and Diyi Yang.
\newblock Swe-smith: Scaling data for software engineering agents, 2025{\natexlab{a}}.
\newblock URL \url{https://arxiv.org/abs/2504.21798}.

\bibitem[Wang et~al.(2025)Wang, Hou, Wei, Tang, and Dong]{wang2025swedev}
Haoran Wang, Zhenyu Hou, Yao Wei, Jie Tang, and Yuxiao Dong.
\newblock Swe-dev: Building software engineering agents with training and inference scaling.
\newblock \url{https://ubecwang.notion.site/1bc32cf963e080b2a01df2895f66021f?v=1bc32cf963e0810ca07e000c86c4c1e1}, 2025.
\newblock Notion Blog.

\bibitem[Aggarwal et~al.(2025)Aggarwal, Kamal, Japesh, Jin, and Schölkopf]{aggarwal2025darsdynamicactionresampling}
Vaibhav Aggarwal, Ojasv Kamal, Abhinav Japesh, Zhijing Jin, and Bernhard Schölkopf.
\newblock Dars: Dynamic action re-sampling to enhance coding agent performance by adaptive tree traversal, 2025.
\newblock URL \url{https://arxiv.org/abs/2503.14269}.

\bibitem[Ma et~al.(2025)Ma, Li, Dong, Jiang, Cao, Chen, Huang, and Li]{ma2025thinkinglongerlargerenhancing}
Yingwei Ma, Yongbin Li, Yihong Dong, Xue Jiang, Rongyu Cao, Jue Chen, Fei Huang, and Binhua Li.
\newblock Thinking longer, not larger: Enhancing software engineering agents via scaling test-time compute, 2025.
\newblock URL \url{https://arxiv.org/abs/2503.23803}.

\bibitem[Wei et~al.(2025)Wei, Duchenne, Copet, Carbonneaux, Zhang, Fried, Synnaeve, Singh, and Wang]{wei2025swerladvancingllmreasoning}
Yuxiang Wei, Olivier Duchenne, Jade Copet, Quentin Carbonneaux, Lingming Zhang, Daniel Fried, Gabriel Synnaeve, Rishabh Singh, and Sida~I. Wang.
\newblock Swe-rl: Advancing llm reasoning via reinforcement learning on open software evolution, 2025.
\newblock URL \url{https://arxiv.org/abs/2502.18449}.

\bibitem[Golubev et~al.(2024)Golubev, Polezhaev, Zainullina, Trofimova, Badertdinov, Anapolskiy, Litvintseva, Karasik, Fisin, Skvortsov, Nekrashevich, Shevtsov, Abramov, and Yangel]{golubev2024search}
Alexander Golubev, Sergey Polezhaev, Karina Zainullina, Maria Trofimova, Ibragim Badertdinov, Yuri Anapolskiy, Daria Litvintseva, Simon Karasik, Filipp Fisin, Sergey Skvortsov, Maxim Nekrashevich, Anton Shevtsov, Sergey Abramov, and Boris Yangel.
\newblock Leveraging training and search for better software engineering agents.
\newblock \emph{Nebius blog}, 2024.
\newblock URL \url{https://nebius.com/blog/posts/training-and-search-for-software-engineering-agents}.

\bibitem[DeepSeek-AI(2024)]{deepseekai2024deepseekv3technicalreport}
DeepSeek-AI.
\newblock Deepseek-v3 technical report, 2024.
\newblock URL \url{https://arxiv.org/abs/2412.19437}.

\bibitem[{Meta AI}(2025)]{meta2025llama4}
{Meta AI}.
\newblock The llama 4 herd: The beginning of a new era of natively multimodal ai innovation, 2025.
\newblock URL \url{https://ai.meta.com/blog/llama-4-multimodal-intelligence/}.

\bibitem[Team(2025)]{qwen3}
Qwen Team.
\newblock Qwen3, April 2025.
\newblock URL \url{https://qwenlm.github.io/blog/qwen3/}.

\bibitem[Lozhkov et~al.(2024)Lozhkov, Li, Allal, Cassano, Lamy-Poirier, Tazi, Tang, Pykhtar, Liu, Wei, Liu, Tian, Kocetkov, Zucker, Belkada, Wang, Liu, Abulkhanov, Paul, Li, Li, Risdal, Li, Zhu, Zhuo, Zheltonozhskii, Dade, Yu, Krauß, Jain, Su, He, Dey, Abati, Chai, Muennighoff, Tang, Oblokulov, Akiki, Marone, Mou, Mishra, Gu, Hui, Dao, Zebaze, Dehaene, Patry, Xu, McAuley, Hu, Scholak, Paquet, Robinson, Anderson, Chapados, Patwary, Tajbakhsh, Jernite, Ferrandis, Zhang, Hughes, Wolf, Guha, von Werra, and de~Vries]{lozhkov2024starcoder2stackv2}
Anton Lozhkov, Raymond Li, Loubna~Ben Allal, Federico Cassano, Joel Lamy-Poirier, Nouamane Tazi, Ao~Tang, Dmytro Pykhtar, Jiawei Liu, Yuxiang Wei, Tianyang Liu, Max Tian, Denis Kocetkov, Arthur Zucker, Younes Belkada, Zijian Wang, Qian Liu, Dmitry Abulkhanov, Indraneil Paul, Zhuang Li, Wen-Ding Li, Megan Risdal, Jia Li, Jian Zhu, Terry~Yue Zhuo, Evgenii Zheltonozhskii, Nii Osae~Osae Dade, Wenhao Yu, Lucas Krauß, Naman Jain, Yixuan Su, Xuanli He, Manan Dey, Edoardo Abati, Yekun Chai, Niklas Muennighoff, Xiangru Tang, Muhtasham Oblokulov, Christopher Akiki, Marc Marone, Chenghao Mou, Mayank Mishra, Alex Gu, Binyuan Hui, Tri Dao, Armel Zebaze, Olivier Dehaene, Nicolas Patry, Canwen Xu, Julian McAuley, Han Hu, Torsten Scholak, Sebastien Paquet, Jennifer Robinson, Carolyn~Jane Anderson, Nicolas Chapados, Mostofa Patwary, Nima Tajbakhsh, Yacine Jernite, Carlos~Muñoz Ferrandis, Lingming Zhang, Sean Hughes, Thomas Wolf, Arjun Guha, Leandro von Werra, and Harm de~Vries.
\newblock Starcoder 2 and the stack v2: The next generation, 2024.
\newblock URL \url{https://arxiv.org/abs/2402.19173}.

\bibitem[Wei et~al.(2024)Wei, Cassano, Liu, Ding, Jain, Mueller, de~Vries, von Werra, Guha, and Zhang]{wei2024selfcodealignselfalignmentcodegeneration}
Yuxiang Wei, Federico Cassano, Jiawei Liu, Yifeng Ding, Naman Jain, Zachary Mueller, Harm de~Vries, Leandro von Werra, Arjun Guha, and Lingming Zhang.
\newblock Selfcodealign: Self-alignment for code generation, 2024.
\newblock URL \url{https://arxiv.org/abs/2410.24198}.

\bibitem[Shao et~al.(2024)Shao, Wang, Zhu, Xu, Song, Bi, Zhang, Zhang, Li, Wu, and Guo]{shao2024deepseekmathpushinglimitsmathematical}
Zhihong Shao, Peiyi Wang, Qihao Zhu, Runxin Xu, Junxiao Song, Xiao Bi, Haowei Zhang, Mingchuan Zhang, Y.~K. Li, Y.~Wu, and Daya Guo.
\newblock Deepseekmath: Pushing the limits of mathematical reasoning in open language models, 2024.
\newblock URL \url{https://arxiv.org/abs/2402.03300}.

\bibitem[Pan et~al.(2024{\natexlab{a}})Pan, Zhang, Tomlin, Zhou, Levine, and Suhr]{pan2024autonomousevaluationrefinementdigital}
Jiayi Pan, Yichi Zhang, Nicholas Tomlin, Yifei Zhou, Sergey Levine, and Alane Suhr.
\newblock Autonomous evaluation and refinement of digital agents, 2024{\natexlab{a}}.
\newblock URL \url{https://arxiv.org/abs/2404.06474}.

\bibitem[Pan et~al.(2024{\natexlab{b}})Pan, Wang, Neubig, Jaitly, Ji, Suhr, and Zhang]{pan2024trainingsoftwareengineeringagents}
Jiayi Pan, Xingyao Wang, Graham Neubig, Navdeep Jaitly, Heng Ji, Alane Suhr, and Yizhe Zhang.
\newblock Training software engineering agents and verifiers with swe-gym, 2024{\natexlab{b}}.
\newblock URL \url{https://arxiv.org/abs/2412.21139}.

\bibitem[Rashid et~al.(2025)Rashid, Bock, Zhuang, Buchholz, Esler, Valentin, Franceschi, Wistuba, Sivaprasad, Kim, Deoras, Zappella, and Callot]{rashid2025swepolybenchmultilanguagebenchmarkrepository}
Muhammad~Shihab Rashid, Christian Bock, Yuan Zhuang, Alexander Buchholz, Tim Esler, Simon Valentin, Luca Franceschi, Martin Wistuba, Prabhu~Teja Sivaprasad, Woo~Jung Kim, Anoop Deoras, Giovanni Zappella, and Laurent Callot.
\newblock Swe-polybench: A multi-language benchmark for repository level evaluation of coding agents, 2025.
\newblock URL \url{https://arxiv.org/abs/2504.08703}.

\bibitem[Badertdinov et~al.(2024)Badertdinov, Trofimova, Anapolskiy, Abramov, Zainullina, Golubev, Polezhaev, Litvintseva, Karasik, Fisin, Skvortsov, Nekrashevich, Shevtsov, and Yangel]{badertdinov2024scaling}
Ibragim Badertdinov, Maria Trofimova, Yuri Anapolskiy, Sergey Abramov, Karina Zainullina, Alexander Golubev, Sergey Polezhaev, Daria Litvintseva, Simon Karasik, Filipp Fisin, Sergey Skvortsov, Maxim Nekrashevich, Anton Shevtsov, and Boris Yangel.
\newblock Scaling data collection for training software engineering agents.
\newblock \emph{Nebius blog}, 2024.
\newblock URL \url{https://nebius.com/blog/posts/scaling-data-collection-for-training-swe-agents}.

\bibitem[{TractoAI}(2025)]{tractoai2025}
{TractoAI}.
\newblock Tractoai: Cloud platform for deploying, scaling, and monitoring ai and big data workloads, 2025.
\newblock URL \url{https://tracto.ai/}.

\bibitem[Grigorik(2011)]{grigorik2011gharchive}
Ilya Grigorik.
\newblock Gh archive, 2011.
\newblock URL \url{https://www.gharchive.org/}.
\newblock Accessed 15 May 2025.

\bibitem[Qwen et~al.(2025)Qwen, :, Yang, Yang, Zhang, Hui, Zheng, Yu, Li, Liu, Huang, Wei, Lin, Yang, Tu, Zhang, Yang, Yang, Zhou, Lin, Dang, Lu, Bao, Yang, Yu, Li, Xue, Zhang, Zhu, Men, Lin, Li, Tang, Xia, Ren, Ren, Fan, Su, Zhang, Wan, Liu, Cui, Zhang, and Qiu]{qwen2025qwen25technicalreport}
Qwen, :, An~Yang, Baosong Yang, Beichen Zhang, Binyuan Hui, Bo~Zheng, Bowen Yu, Chengyuan Li, Dayiheng Liu, Fei Huang, Haoran Wei, Huan Lin, Jian Yang, Jianhong Tu, Jianwei Zhang, Jianxin Yang, Jiaxi Yang, Jingren Zhou, Junyang Lin, Kai Dang, Keming Lu, Keqin Bao, Kexin Yang, Le~Yu, Mei Li, Mingfeng Xue, Pei Zhang, Qin Zhu, Rui Men, Runji Lin, Tianhao Li, Tianyi Tang, Tingyu Xia, Xingzhang Ren, Xuancheng Ren, Yang Fan, Yang Su, Yichang Zhang, Yu~Wan, Yuqiong Liu, Zeyu Cui, Zhenru Zhang, and Zihan Qiu.
\newblock Qwen2.5 technical report, 2025.
\newblock URL \url{https://arxiv.org/abs/2412.15115}.

\bibitem[Yao et~al.(2023)Yao, Zhao, Yu, Du, Shafran, Narasimhan, and Cao]{yao2023reactsynergizingreasoningacting}
Shunyu Yao, Jeffrey Zhao, Dian Yu, Nan Du, Izhak Shafran, Karthik Narasimhan, and Yuan Cao.
\newblock React: Synergizing reasoning and acting in language models, 2023.
\newblock URL \url{https://arxiv.org/abs/2210.03629}.

\bibitem[Chen et~al.(2021)Chen, Tworek, Jun, Yuan, de~Oliveira~Pinto, Kaplan, Edwards, Burda, Joseph, Brockman, Ray, Puri, Krueger, Petrov, Khlaaf, Sastry, Mishkin, Chan, Gray, Ryder, Pavlov, Power, Kaiser, Bavarian, Winter, Tillet, Such, Cummings, Plappert, Chantzis, Barnes, Herbert-Voss, Guss, Nichol, Paino, Tezak, Tang, Babuschkin, Balaji, Jain, Saunders, Hesse, Carr, Leike, Achiam, Misra, Morikawa, Radford, Knight, Brundage, Murati, Mayer, Welinder, McGrew, Amodei, McCandlish, Sutskever, and Zaremba]{chen2021evaluatinglargelanguagemodels}
Mark Chen, Jerry Tworek, Heewoo Jun, Qiming Yuan, Henrique~Ponde de~Oliveira~Pinto, Jared Kaplan, Harri Edwards, Yuri Burda, Nicholas Joseph, Greg Brockman, Alex Ray, Raul Puri, Gretchen Krueger, Michael Petrov, Heidy Khlaaf, Girish Sastry, Pamela Mishkin, Brooke Chan, Scott Gray, Nick Ryder, Mikhail Pavlov, Alethea Power, Lukasz Kaiser, Mohammad Bavarian, Clemens Winter, Philippe Tillet, Felipe~Petroski Such, Dave Cummings, Matthias Plappert, Fotios Chantzis, Elizabeth Barnes, Ariel Herbert-Voss, William~Hebgen Guss, Alex Nichol, Alex Paino, Nikolas Tezak, Jie Tang, Igor Babuschkin, Suchir Balaji, Shantanu Jain, William Saunders, Christopher Hesse, Andrew~N. Carr, Jan Leike, Josh Achiam, Vedant Misra, Evan Morikawa, Alec Radford, Matthew Knight, Miles Brundage, Mira Murati, Katie Mayer, Peter Welinder, Bob McGrew, Dario Amodei, Sam McCandlish, Ilya Sutskever, and Wojciech Zaremba.
\newblock Evaluating large language models trained on code, 2021.
\newblock URL \url{https://arxiv.org/abs/2107.03374}.

\bibitem[Austin et~al.(2021)Austin, Odena, Nye, Bosma, Michalewski, Dohan, Jiang, Cai, Terry, Le, and Sutton]{austin2021programsynthesislargelanguage}
Jacob Austin, Augustus Odena, Maxwell Nye, Maarten Bosma, Henryk Michalewski, David Dohan, Ellen Jiang, Carrie Cai, Michael Terry, Quoc Le, and Charles Sutton.
\newblock Program synthesis with large language models, 2021.
\newblock URL \url{https://arxiv.org/abs/2108.07732}.

\bibitem[Hendrycks et~al.(2021)Hendrycks, Basart, Kadavath, Mazeika, Arora, Guo, Burns, Puranik, He, Song, and Steinhardt]{hendrycks2021measuringcodingchallengecompetence}
Dan Hendrycks, Steven Basart, Saurav Kadavath, Mantas Mazeika, Akul Arora, Ethan Guo, Collin Burns, Samir Puranik, Horace He, Dawn Song, and Jacob Steinhardt.
\newblock Measuring coding challenge competence with apps, 2021.
\newblock URL \url{https://arxiv.org/abs/2105.09938}.

\bibitem[Jain et~al.(2024)Jain, Han, Gu, Li, Yan, Zhang, Wang, Solar-Lezama, Sen, and Stoica]{jain2024livecodebenchholisticcontaminationfree}
Naman Jain, King Han, Alex Gu, Wen-Ding Li, Fanjia Yan, Tianjun Zhang, Sida Wang, Armando Solar-Lezama, Koushik Sen, and Ion Stoica.
\newblock Livecodebench: Holistic and contamination free evaluation of large language models for code, 2024.
\newblock URL \url{https://arxiv.org/abs/2403.07974}.

\bibitem[Aleithan et~al.(2024)Aleithan, Xue, Mohajer, Nnorom, Uddin, and Wang]{aleithan2024swebenchenhancedcodingbenchmark}
Reem Aleithan, Haoran Xue, Mohammad~Mahdi Mohajer, Elijah Nnorom, Gias Uddin, and Song Wang.
\newblock Swe-bench+: Enhanced coding benchmark for llms, 2024.
\newblock URL \url{https://arxiv.org/abs/2410.06992}.

\bibitem[Zan et~al.(2025)Zan, Huang, Liu, Chen, Zhang, Xin, Chen, Liu, Zhong, Li, Liu, Xiao, Chen, Zhang, Su, Liu, Long, Shen, and Xiang]{zan2025multiswebenchmultilingualbenchmarkissue}
Daoguang Zan, Zhirong Huang, Wei Liu, Hanwu Chen, Linhao Zhang, Shulin Xin, Lu~Chen, Qi~Liu, Xiaojian Zhong, Aoyan Li, Siyao Liu, Yongsheng Xiao, Liangqiang Chen, Yuyu Zhang, Jing Su, Tianyu Liu, Rui Long, Kai Shen, and Liang Xiang.
\newblock Multi-swe-bench: A multilingual benchmark for issue resolving, 2025.
\newblock URL \url{https://arxiv.org/abs/2504.02605}.

\bibitem[Muennighoff et~al.(2024)Muennighoff, Liu, Zebaze, Zheng, Hui, Zhuo, Singh, Tang, von Werra, and Longpre]{muennighoff2024octopackinstructiontuningcode}
Niklas Muennighoff, Qian Liu, Armel Zebaze, Qinkai Zheng, Binyuan Hui, Terry~Yue Zhuo, Swayam Singh, Xiangru Tang, Leandro von Werra, and Shayne Longpre.
\newblock Octopack: Instruction tuning code large language models, 2024.
\newblock URL \url{https://arxiv.org/abs/2308.07124}.

\bibitem[Luo et~al.(2023)Luo, Xu, Zhao, Sun, Geng, Hu, Tao, Ma, Lin, and Jiang]{luo2023wizardcoderempoweringcodelarge}
Ziyang Luo, Can Xu, Pu~Zhao, Qingfeng Sun, Xiubo Geng, Wenxiang Hu, Chongyang Tao, Jing Ma, Qingwei Lin, and Daxin Jiang.
\newblock Wizardcoder: Empowering code large language models with evol-instruct, 2023.
\newblock URL \url{https://arxiv.org/abs/2306.08568}.

\bibitem[Liu et~al.(2024)Liu, Nguyen, Shang, Ding, Li, Yu, Kumar, and Wang]{liu2024learningcodepreferencesynthetic}
Jiawei Liu, Thanh Nguyen, Mingyue Shang, Hantian Ding, Xiaopeng Li, Yu~Yu, Varun Kumar, and Zijian Wang.
\newblock Learning code preference via synthetic evolution, 2024.
\newblock URL \url{https://arxiv.org/abs/2410.03837}.

\bibitem[DeepSeek-AI(2025)]{deepseekai2025deepseekr1incentivizingreasoningcapability}
DeepSeek-AI.
\newblock Deepseek-r1: Incentivizing reasoning capability in llms via reinforcement learning, 2025.
\newblock URL \url{https://arxiv.org/abs/2501.12948}.

\bibitem[Seed et~al.(2025)Seed, :, Chen, Fan, Liu, Liu, Lin, Wang, Wang, Wei, Xu, Yuan, Yue, Yan, Yu, Zuo, Zhang, Zhu, An, Bai, Bao, Bin, Chen, Chen, Chen, Chen, Chen, Chen, Chen, Chen, Chen, Chen, Chen, Chen, Chi, Dai, Dai, Dai, Dou, Du, Du, Duan, Dun, Fan, Feng, Feng, Feng, Fu, Fu, Fu, Ge, Guo, Han, Han, Hao, Hao, He, He, He, Heng, Hong, Hou, Hu, Hu, Hu, Hua, Huang, Huang, Huang, Huang, Huang, Huang, Jia, Jia, Jia, Jiang, Jiang, Jiang, Jiang, Jiang, Jiao, Jin, Jin, Lai, Li, Li, Li, Li, Li, Wan, Wang, Li, Li, Li, Li, Li, Li, Li, Li, Liang, Liang, Lin, Lin, Lin, Liu, Liu, Liu, Liu, Liu, Liu, Liu, Liu, Liu, Liu, Liu, Liu, Liu, Liu, Liu, Liu, Long, Lou, Lou, Luo, Luo, Lv, Lv, Ma, Ma, Ma, Ma, Ma, Ma, Ma, Mao, Min, Nan, Ning, Ou, Pan, Pang, Peng, Peng, Qian, Qian, Qiao, Qu, Ren, Ren, Shan, Shen, Shen, Shen, Sheng, Shi, Shi, Shi, Cao, Song, Song, Su, Sun, Sun, Sun, Wan, Wang, Wang, Wang, Wang, Wang, Wang, Wang, Wang, Wang, Wang, Wang, Wang, Wang, Wang, Wang, Wang, Wang, Wei, Wei, Wei, Wei, Wu, Wu, Wu, Wu, Wu, Wu,
  Wu, Wu, Wu, Xi, Xia, Xian, Xiang, Xiang, Xiao, Xiao, Xiao, Xiao, Xin, Xin, Xiong, Xu, Xu, Xu, Xu, Xu, Xu, Xu, Xu, Yan, Yan, Yang, Yang, Yang, Yang, Yang, Yang, Yang, Yang, Yang, Yao, Yi, Yin, Yin, Ying, Yu, Yu, Yu, Yu, Yu, Yuan, Yuan, Zeng, Zhan, Zhang, Zhang, Zhang, Zhang, Zhang, Zhang, Zhang, Zhang, Zhang, Zhang, Zhang, Zhang, Zhang, Zhang, Zhang, Zhang, Zhang, Zheng, Zheng, Zheng, Zheng, Zheng, Zhi, Zhong, Zhong, Zhong, Zhong, Zhou, Zhou, Zhou, Zhu, Zhu, Zhu, and Zuo]{seed2025seed15thinkingadvancingsuperbreasoning}
ByteDance Seed, :, Jiaze Chen, Tiantian Fan, Xin Liu, Lingjun Liu, Zhiqi Lin, Mingxuan Wang, Chengyi Wang, Xiangpeng Wei, Wenyuan Xu, Yufeng Yuan, Yu~Yue, Lin Yan, Qiying Yu, Xiaochen Zuo, Chi Zhang, Ruofei Zhu, Zhecheng An, Zhihao Bai, Yu~Bao, Xingyan Bin, Jiangjie Chen, Feng Chen, Hongmin Chen, Riwei Chen, Liangqiang Chen, Zixin Chen, Jinsong Chen, Siyan Chen, Kaiyuan Chen, Zhi Chen, Jin Chen, Jiecao Chen, Jinxin Chi, Weinan Dai, Ning Dai, Jiahui Dai, Shihan Dou, Yantao Du, Zhengyin Du, Jianhui Duan, Chen Dun, Ting-Han Fan, Jiazhan Feng, Junda Feng, Ziyuan Feng, Yuwei Fu, Wenqi Fu, Hanjie Fu, Hao Ge, Hongyi Guo, Mingji Han, Li~Han, Wenhao Hao, Xintong Hao, Qianyu He, Jerry He, Feng He, Wen Heng, Zehua Hong, Qi~Hou, Liang Hu, Shengding Hu, Nan Hu, Kai Hua, Qi~Huang, Ziyue Huang, Hongzhi Huang, Zihao Huang, Ting Huang, Wenhao Huang, Wei Jia, Bin Jia, Xiaoying Jia, Yuhua Jiang, Haobin Jiang, Ziheng Jiang, Kaihua Jiang, Chengquan Jiang, Jianpeng Jiao, Xiaoran Jin, Xing Jin, Xunhao Lai, Zheng Li, Xiang Li, Liyi
  Li, Hongkai Li, Zheng Li, Shengxian Wan, Ya~Wang, Yunshui Li, Chenggang Li, Niuniu Li, Siyu Li, Xi~Li, Xiao Li, Aoyan Li, Yuntao Li, Nianning Liang, Xinnian Liang, Haibin Lin, Weijian Lin, Ye~Lin, Zhicheng Liu, Guanlin Liu, Guanlin Liu, Chenxiao Liu, Yan Liu, Gaohong Liu, Juncai Liu, Chundian Liu, Deyi Liu, Kaibo Liu, Siyao Liu, Qi~Liu, Yongfei Liu, Kang Liu, Gan Liu, Boyi Liu, Rui Long, Weiqiang Lou, Chenwei Lou, Xiang Luo, Yao Luo, Caiping Lv, Heyang Lv, Bole Ma, Qianli Ma, Hongzhi Ma, Yiyuan Ma, Jin Ma, Wenchang Ma, Tingting Ma, Chen Mao, Qiyang Min, Zhe Nan, Guanghan Ning, Jinxiang Ou, Haojie Pan, Renming Pang, Yanghua Peng, Tao Peng, Lihua Qian, Lihua Qian, Mu~Qiao, Meng Qu, Cheng Ren, Hongbin Ren, Yong Shan, Wei Shen, Ke~Shen, Kai Shen, Guangming Sheng, Jinlong Shi, Wenlei Shi, Guang Shi, Shuai~Shuai Cao, Yuxin Song, Zuquan Song, Jing Su, Yifan Sun, Tao Sun, Zewei Sun, Borui Wan, Zihan Wang, Xiaohui Wang, Xi~Wang, Shuguang Wang, Jun Wang, Qinlong Wang, Chenyuan Wang, Shuai Wang, Zihan Wang, Changbao
  Wang, Jiaqiang Wang, Shihang Wang, Xuwu Wang, Zaiyuan Wang, Yuxuan Wang, Wenqi Wang, Taiqing Wang, Chengzhi Wei, Houmin Wei, Ziyun Wei, Shufa Wei, Zheng Wu, Yonghui Wu, Yangjun Wu, Bohong Wu, Shuang Wu, Jingqiao Wu, Ning Wu, Shuangzhi Wu, Jianmin Wu, Chenguang Xi, Fan Xia, Yuqiao Xian, Liang Xiang, Boren Xiang, Bowen Xiao, Zhen Xiao, Xia Xiao, Yongsheng Xiao, Chao Xin, Shulin Xin, Yuwen Xiong, Jingjing Xu, Ziwen Xu, Chenyin Xu, Jiayi Xu, Yifan Xu, Wei Xu, Yufei Xu, Shikun Xu, Shipeng Yan, Shen Yan, Qingping Yang, Xi~Yang, Tianhao Yang, Yuehang Yang, Yuan Yang, Ximing Yang, Zeyu Yang, Guang Yang, Yifan Yang, Xuesong Yao, Bairen Yi, Fan Yin, Jianian Yin, Ziqiang Ying, Xiangyu Yu, Hongli Yu, Song Yu, Menghan Yu, Huan Yu, Siyu Yuan, Jun Yuan, Yutao Zeng, Tianyang Zhan, Zheng Zhang, Yun Zhang, Mofan Zhang, Wang Zhang, Ru~Zhang, Zhi Zhang, Tianqi Zhang, Xinyi Zhang, Zhexi Zhang, Sijun Zhang, Wenqiang Zhang, Xiangxiang Zhang, Yongtao Zhang, Yuyu Zhang, Ge~Zhang, He~Zhang, Yue Zhang, Renjie Zheng, Ningxin Zheng,
  Zhuolin Zheng, Yaowei Zheng, Chen Zheng, Xiaoyun Zhi, Wanjun Zhong, Cheng Zhong, Zheng Zhong, Baoquan Zhong, Xun Zhou, Na~Zhou, Huan Zhou, Hang Zhu, Defa Zhu, Wenjia Zhu, and Lei Zuo.
\newblock Seed1.5-thinking: Advancing superb reasoning models with reinforcement learning, 2025.
\newblock URL \url{https://arxiv.org/abs/2504.13914}.

\bibitem[OpenThoughts(2025)]{openthoughts}
Team OpenThoughts.
\newblock {Open Thoughts}, January 2025.
\newblock URL \url{https://open-thoughts.ai}.

\bibitem[Xie et~al.(2025)Xie, Xie, Sheth, Liu, Fried, and Rose]{xie2025repostscalablerepositorylevelcoding}
Yiqing Xie, Alex Xie, Divyanshu Sheth, Pengfei Liu, Daniel Fried, and Carolyn Rose.
\newblock Repost: Scalable repository-level coding environment construction with sandbox testing, 2025.
\newblock URL \url{https://arxiv.org/abs/2503.07358}.

\bibitem[Luo et~al.(2025)Luo, Jain, Singh, Tan, Patel, Wu, Ariyak, Cai, Venkat, Zhu, Athiwaratkun, Roongta, Zhang, Li, Popa, Sen, and Stoica]{deepswe2025}
Michael Luo, Naman Jain, Jaskirat Singh, Sijun Tan, Ameen Patel, Qingyang Wu, Alpay Ariyak, Colin Cai, Tarun Venkat, Shang Zhu, Ben Athiwaratkun, Manan Roongta, Ce~Zhang, Li~Erran Li, Raluca~Ada Popa, Koushik Sen, and Ion Stoica.
\newblock Deepswe: Training a state-of-the-art coding agent from scratch by scaling rl.
\newblock \url{https://pretty-radio-b75.notion.site/DeepSWE-Training-a-Fully-Open-sourced-State-of-the-Art-Coding-Agent-by-Scaling-RL-22281902c1468193aabbe9a8c59bbe33}, 2025.
\newblock Notion Blog.

\bibitem[Cao et~al.(2025)Cao, Hegde, Li, Griggs, Liu, Tang, Pan, Wang, Malik, Neubig, Hakhamaneshi, Liaw, Moritz, Zaharia, Gonzalez, and Stoica]{cao2025skyrl}
Shiyi Cao, Sumanth Hegde, Dacheng Li, Tyler Griggs, Shu Liu, Eric Tang, Jiayi Pan, Xingyao Wang, Akshay Malik, Graham Neubig, Kourosh Hakhamaneshi, Richard Liaw, Philipp Moritz, Matei Zaharia, Joseph~E. Gonzalez, and Ion Stoica.
\newblock Skyrl-v0: Train real-world long-horizon agents via reinforcement learning, 2025.

\bibitem[Yang et~al.(2025{\natexlab{b}})Yang, Wang, Fu, He, Xiong, Liu, Miao, Gao, Wang, Ma, Li, Liu, Hu, Zhang, Wang, Chen, Sung, Liu, Gao, Yang, and Liu]{yang2025kimidevagentlesstrainingskill}
Zonghan Yang, Shengjie Wang, Kelin Fu, Wenyang He, Weimin Xiong, Yibo Liu, Yibo Miao, Bofei Gao, Yejie Wang, Yingwei Ma, Yanhao Li, Yue Liu, Zhenxing Hu, Kaitai Zhang, Shuyi Wang, Huarong Chen, Flood Sung, Yang Liu, Yang Gao, Zhilin Yang, and Tianyu Liu.
\newblock Kimi-dev: Agentless training as skill prior for swe-agents, 2025{\natexlab{b}}.
\newblock URL \url{https://arxiv.org/abs/2509.23045}.

\bibitem[Kwon et~al.(2023)Kwon, Li, Zhuang, Sheng, Zheng, Yu, Gonzalez, Zhang, and Stoica]{kwon2023efficient}
Woosuk Kwon, Zhuohan Li, Siyuan Zhuang, Ying Sheng, Lianmin Zheng, Cody~Hao Yu, Joseph~E. Gonzalez, Hao Zhang, and Ion Stoica.
\newblock Efficient memory management for large language model serving with pagedattention.
\newblock In \emph{Proceedings of the ACM SIGOPS 29th Symposium on Operating Systems Principles}, 2023.

\end{thebibliography}

\newpage




\appendix

\lstset{
  basicstyle=\ttfamily,
  breaklines=true,
  frame=single,
  backgroundcolor=\color{gray!10},
  columns=fullflexible
}

\section{Related work}

\paragraph{Benchmarks for software engineering} With the progress of LLM capabilities, there has been a growing need for benchmarks to evaluating their coding performance. While HumanEval \citep{chen2021evaluatinglargelanguagemodels}, MBPP \citep{austin2021programsynthesislargelanguage}, and APPS \citep{hendrycks2021measuringcodingchallengecompetence} have proven valuable for standalone function-level evaluation, their utility has largely saturated for state-of-the-art models. Consequently, LiveCodeBench \citep{jain2024livecodebenchholisticcontaminationfree} has emerged, which is continuously updated to reflect the evolving capabilities of LLMs. However, these benchmarks often differ significantly from how LLMs are applied in real-world development workflows, which typically involve repository-level context and interaction. SWE-bench \citep{jimenez2024swebench} introduced execution-based, repository-level tasks sourced from real-world repositories, offering a more realistic evaluation paradigm. A challenge with SWE-bench is that many new LLMs were released after its construction, raising concerns about its continued relevance and potential contamination. Several follow-up works, such as SWE-bench+ \citep{aleithan2024swebenchenhancedcodingbenchmark} and Multi-SWE-bench \citep{zan2025multiswebenchmultilingualbenchmarkissue}, address specific shortcomings like post-cutoff filtering and multilingual annotation. Nevertheless, all these extensions fundamentally rely on a manually curated construction process. This manual element complicates timely updates and makes it difficult to keep the benchmark consistently aligned with the rapidly advancing capabilities of LLMs. In contrast, SWE-rebench automates the entire process of task creation from real-world repositories at scale. This automation enables a new class of dataset and benchmark: one that is task-rich, diverse, regularly updated, and suitable for both training and evaluation without the bottlenecks of manual curation.

\paragraph{Training datasets for code generation} Most prior work on collecting data to improve LLMs for code has focused on supervised fine-tuning \citep{muennighoff2024octopackinstructiontuningcode, luo2023wizardcoderempoweringcodelarge, liu2024learningcodepreferencesynthetic}, often by leveraging samples from more powerful proprietary models. On the other hand, the remarkable success of large-scale reinforcement learning (RL) in math reasoning tasks \citep{deepseekai2025deepseekr1incentivizingreasoningcapability, seed2025seed15thinkingadvancingsuperbreasoning, openthoughts} implies that it should also excel in software engineering, as both domains involve rich, multi-step workflows with potentially verifiable outcomes. However, leveraging~RL effectively in this setting demands gathering a large collection of real-world, interactive SWE tasks which is challenging, as each task requires the construction of a suitable and reliable execution environment. Some works attempt to gather such tasks by artificially injecting bugs into code \citep{yang2025swesmithscalingdatasoftware} or generating synthetic tests \citep{xie2025repostscalablerepositorylevelcoding}. While these approaches can generate data at scale, they may not fully capture the nuances and complexities of real-world software issues. The original SWE-bench dataset and its extensions for training \citep{pan2024trainingsoftwareengineeringagents} provide valuable real-world tasks, but still suffer from limited diversity and scale due to their reliance on manual construction. In contrast, SWE-rebench reflects real-world development processes and preserves high task quality through its grounding in real repositories, while achieving significantly greater scale and diversity via its automated construction pipeline. This makes it particularly well-suited for training agents through an automated construction pipeline. Indeed, the value of this approach is demonstrated by research successfully applying reinforcement learning to train software engineering agents \citep{deepswe2025, cao2025skyrl}, including work that directly utilizes our SWE-rebench dataset \citep{yang2025kimidevagentlesstrainingskill}.

\section{Automated dependency installation}

As a reminder, the goal of automated dependency installation is to generate correct environment setup instructions for each task instance without using manual curation. The process is entirely LLM-driven and consists of the following sub-stages:

\begin{enumerate}
    \item \textbf{Identifying relevant files:} An LLM is asked to analyze the list of repository files and select those most likely to contain installation instructions, such as \texttt{README.md}, \texttt{setup.py}, or \texttt{requirements.txt}.
    \item \textbf{Extracting installation recipe:} Based on the content of files found in the previous stage, the model generates a structured JSON recipe specifying how to install dependencies and run tests for the repository.
    \item \textbf{Updating the recipe (if needed):} If the recipe fails, another LLM call is made to analyze the failure and refine the recipe.
\end{enumerate}

In the following subsections, we show the exact prompts used at each of these stages, along with examples of LLM-generated outputs. These materials are intended to provide transparency and facilitate reproducibility of our environment setup pipeline.

\subsection{Identifying installation-related files}
\label{sec:identify_files}

The automated generation of installation recipes, a key part of our agentless approach described in Section~\ref{sec:automated_install_conf}, begins with identifying repository files that are likely to contain setup instructions. To achieve this, an LLM is provided with a list of all files present in the repository. The LLM's task is to analyze this list and return a curated subset of file paths deemed most relevant for extracting installation, dependency, and testing information. The prompt below details the instructions given to the LLM for this file identification step. The output of this step (a JSON array of file paths) subsequently serves as input for the next LLM call, which extracts the actual installation commands.

\begin{lstlisting}[upquote=true,caption=LLM prompt for identifying installation-related files]

You are tasked with identifying files that likely contain installation instructions for a GitHub repository.

Repository: {{ repo_name }}

Below is a list of files in the repository that may be helpful for understanding the installation and setup process:

{{ list_of_files }}

Please analyze this list and identify the files that are most likely to contain information about:
  * Installation instructions
  * Dependencies or requirements
  * Setup procedures
  * Development environment configuration
  * Testing setup

Think step by step:
  * Identify README files, as they often contain installation instructions.
  * Look for setup.py, pyproject.toml, requirements.txt, environment.yml.
  * Consider files in directories like \texttt{docs} that might contain installation guides.
  * Look for test configuration files that might help understand how to run tests.
  * Consider \textbf{only} files from the list provided above.
  * Prioritize files in the repository root (top-level directory).
  * Only include files from subdirectories if they are clearly relevant to installation or setup.

Return **only** a JSON array containing the paths to the most relevant files for installation and setup. Include only files that are directly relevant to the tasks above. Sort the files from most to least relevant and limit your response to no more than 10 files, preferring fewer files that are truly essential.

For example:
[
    "README.md",
    "setup.py",
    "requirements.txt"
]
\end{lstlisting}

\subsection{Extracting installation recipe}

Following the identification of relevant files, the next step in our automated pipeline (Section~\ref{sec:automated_install_conf}) is to extract a concrete installation recipe. For this, the LLM receives the concatenated content of the previously selected relevant files. The primary task of the LLM is to analyze this content and synthesize a structured JSON object, \textit{installation recipe}, which encapsulates all necessary commands and configurations to set up the project environment, install dependencies, and execute tests. The prompt presented below guides the LLM through this extraction process, emphasizing the expected JSON structure and providing context on how the generated recipe will be utilized by downstream automation scripts.

\label{sec:extract_install_recipe}

\begin{lstlisting}[upquote=true,caption=LLM prompt for extracting installation recipe]
You are tasked with extracting detailed installation instructions from the following repository files. Repository: {{ repo_name }}

Please analyze the content of these files and extract comprehensive installation instructions: {{ rendered[:100000] }}

First, think step by step. After your analysis, return your findings in the following JSON format:

{
  "python": "3.9",
  "packages": "requirements.txt",
  "install": "pip install -e .[dev]",
  "test_cmd": "pytest --no-header -rA --tb=line --color=no -p no:cacheprovider -W ignore::DeprecationWarning",
  "pre_install": ["apt-get update", "apt-get install -y gcc"],
  "reqs_path": ["requirements/base.txt"],
  "env_yml_path": ["environment.yml"],
  "pip_packages": ["numpy>=1.16.0", "pandas>=1.0.0"]
}


Here is how this JSON will be used:

```bash
git clone <repo_url> repo
cd repo
git checkout <base_sha>
bash <pre_install>
conda create -n <repo> python=<python> -y
conda activate <repo>
if <packages> == requirements.txt; then
  for path in <reqs_path>:
      pip install -r $path
elif <packages> == environment.yml; then
  for path in <env_yml_path>:
      conda env update -f $path
else:
  pip install <packages>
pip install <pip_packages>
bash <install>
bash <test_cmd>
```

**IMPORTANT:**
  * For the "install" field, always use local install commands like pip install -e .[dev]
  * Do NOT include packages in pip_packages that will be installed by pip install -e .
  * Include only explicitly needed packages in pip_packages.
  * reqs_path and env_yml_path must match filenames from the provided files (e.g., [File: filename]).
  * If "packages" is requirements.txt, you must provide at least one reqs_path.
  * Add relevant test frameworks to pip_packages (e.g., pytest, nose).
  * Use -y in all conda commands.
  * Prefer direct and specific pytest commands over general wrappers.
  * Avoid test commands with placeholders like {test_name}.
  * If a Makefile runs tests, extract the actual test command (e.g., pytest).

You must ensure the final JSON includes required fields (, install, test_cmd), and optionally packages, pre_install, reqs_path, env_yml_path, pip_packages if relevant.

Base your reasoning on all provided files and return it followed by the JSON output in a Markdown code block (without comments).
    
\end{lstlisting}

\subsection{Example of LLM-generated installation recipe}
\label{appendix:json_recipe_example}

Below is an illustrative example of an output from the installation recipe extraction stage that includes both the reasoning and the final JSON.

\begin{lstlisting}[upquote=true,caption=LLM-generated installation recipe]
1. Python version. The setup.py file specifies that the project requires Python 3.6 or greater. The README.rst and setup.py files also indicate that the project supports Python 3.6, 3.7, 3.8, 3.9, and 3.10. For a safe middle-ground, Python 3.9 is chosen. 

2. Installation Method. The README.rst and DEVELOPMENT.md files both mention using pip to install the project. Specifically, the development installation command is pip install -e .[dev]

3. Dependencies. The setup.py file lists flake8 >= 3.0.0 and attrs >= 19.2.0 as dependencies. The DEVELOPMENT.md file also mentions additional development dependencies: coverage, hypothesis, hypothesmith >= 0.2, and pre-commit. These are included in the pip_packages field.

4. Test command. The README.rst and DEVELOPMENT.md files both mention running tests with coverage run tests/test_bugbear.py. However, for simplicity and to follow the guidelines, the test command is simplified to pytest tests/test_bugbear.py.

5. Pre-install steps. No system-level pre-install steps are mentioned in the provided files.

6. Additional packages. The pytest package is needed for running tests, and it is not explicitly listed in the project's dependencies. Therefore, it is added to the pip_packages field.

{
  "python": "3.9",
  "install": "pip install -e .[dev]",
  "test_cmd": "pytest tests/test_bugbear.py",
  "pip_packages": [
    "pytest",
    "coverage",
    "hypothesis",
    "hypothesmith>=0.2",
    "pre-commit"
  ]
}   
\end{lstlisting}

\subsection{Updating installation recipe}

As described in Section~\ref{sec:automated_install_conf}, our agentless installation recipe generation includes an iterative refinement step. If the initial attempt to set up an environment using an LLM-generated recipe fails, we trigger a correction mechanism. This mechanism involves a subsequent call to the LLM, providing it with the original (failed) installation recipe, the error logs produced during the failed setup or test execution, as well as the contents of relevant files. The LLM's objective is then to diagnose the errors based on the logs and propose modifications to the recipe to resolve these issues. The prompt detailed below instructs the LLM on how to perform this error analysis and generate an updated, corrected installation recipe in the same structured JSON format.

\label{sec:update_install_recipe}

\begin{lstlisting}[upquote=true,caption=LLM prompt for updating installation recipe]
You are an expert in fixing software installation and testing issues. Analyze the installation logs provided and update the installation configuration to fix any errors.

Install config fields:

{
  "python": "3.9",
  "packages": "requirements.txt",
  "install": "pip install -e .[dev]",
  "test_cmd": "pytest --no-header -rA --tb=line --color=no -p no:cacheprovider -W ignore::DeprecationWarning",
  "pre_install": ["apt-get update", "apt-get install -y gcc"],
  "reqs_path": ["requirements/base.txt"],
  "env_yml_path": ["environment.yml"],
  "pip_packages": ["numpy>=1.16.0", "pandas>=1.0.0"]
}

Current installation configuration:
{{ install_config }}

Error logs from installation/testing (last relevant lines): {{ cut_logs }}

Your task:
  * Identify the root causes of the errors in the logs.
  * Modify the installation configuration to address these issues.
  * You might need to:
    - Add missing dependencies
    - Fix command syntax errors
    - Change installation order
    - Add environment variables
    - Modify test commands

First, provide brief reasoning (<100 words) about what's causing the errors and your planned fixes.

Then return the complete updated installation configuration as a valid JSON object.
\end{lstlisting}

\subsection{Failure modes during execution validation and mitigation strategies}
\label{appendix:failure_modes}

During execution validation, we identified several recurring failure modes that prevent otherwise valid tasks from being included in the dataset:

\begin{itemize}
    \item \textbf{Flaky tests with external API calls or floating-point precision issues:} Many failures were attributed to tests involving external API calls to intermittently unavailable endpoints or tests asserting exact equality of floating-point numbers. These failures are non-deterministic and do not reliably reflect task quality. Since we cannot depend on such tests for consistent verification, we run tests multiple times and exclude instances that fail in at least one run.
    
    \item \textbf{Lack of interactivity during installation:} Some projects, particularly older ones, require an interactive approach during the installation process, where dependencies must be incrementally installed or configurations adjusted based on intermediate error messages. Such iterative troubleshooting cannot be reliably accomplished by an agent with a small number of predefined steps, leading to installation failures even when the underlying task is valid.
    
    \item \textbf{Unpinned and obsolete dependencies:} Environment setup frequently failed for older repositories where README instructions were accurate but dependency versions were not pinned or were no longer publicly available. To mitigate this, we freeze all dependency versions after successful setup and maintain internal package mirrors for common dependencies, significantly reducing setup failures on retry.
\end{itemize}

These failure modes account for most validation rejections, and addressing them---particularly through more interactive installation mechanisms or advanced dependency resolution---represents a key opportunity for improving the pipeline's yield.

\section{Evaluation of automated installation recipe generation}
\label{appendix:install_method_eval}

To assess the efficacy of our automated installation recipe generation and to select effective prompts, we conducted a validation study. We utilized a curated set of 18 task instances, each from a distinct repository within the original SWE-bench dataset. These instances were chosen because their environment installation instructions were originally manually crafted, providing a reliable baseline. For each instance, we attempted to automatically generate an installation recipe, set up the corresponding environment, and execute its tests.

The performance of the agentless LLM-based recipe generation (with varying numbers of candidate generations) was compared against an interactive agent designed for the same purpose. The results are summarized in Table~\ref{tab:install_method_comparison}. Some unsuccessful configuration attempts are due to limitations in our automated log parsing during verification, especially for repositories with custom test frameworks such as Django. Our generic parsers (e.g., for pytest) may fail to extract test results from highly customised test outputs. Nevertheless, this small-scale validation served as a sanity check and informed our selection of prompts and the number of candidate generations for the agentless approach in the main pipeline. The list of specific SWE-bench instances used for this validation can be found in Appendix~\ref{list_of_instances}.

\begin{table}[htbp]
  \centering
  \caption{Comparison of installation method success rates on 18 SWE-bench tasks.}
  \label{tab:install_method_comparison}
  \begin{tabular}{llcc}
    \toprule
    Approach & Candidates/Trials & Successfully Configured \\
    \midrule
    Agentless & 1 & 6 out of 18 \\
    Agentless & 3 & 8 out of 18 \\
    Agentless & 10 & 9 out of 18 \\
    Agent-based & 1 & 8 out of 18 \\
    \bottomrule
  \end{tabular}
\end{table}

\subsection{List of instances to validate automatic installation instruction extraction}
\label{list_of_instances}

\begin{itemize}
    \item \texttt{astropy\_\_astropy-12907}
    \item \texttt{django\_\_django-15315}
    \item \texttt{marshmallow-code\_\_marshmallow-1164}
    \item \texttt{matplotlib\_\_matplotlib-20826}
    \item \texttt{mwaskom\_\_seaborn-3069}
    \item \texttt{pallets\_\_flask-5014}
    \item \texttt{psf\_\_requests-2317}
    \item \texttt{pvlib\_\_pvlib-python-1160}
    \item \texttt{pydata\_\_xarray-6744}
    \item \texttt{pydicom\_\_pydicom-996}
    \item \texttt{pylint-dev\_\_astroid-1741}
    \item \texttt{pylint-dev\_\_pylint-4604}
    \item \texttt{pytest-dev\_\_pytest-7982}
    \item \texttt{pyvista\_\_pyvista-4853}
    \item \texttt{scikit-learn\_\_scikit-learn-13142}
    \item \texttt{sphinx-doc\_\_sphinx-8721}
    \item \texttt{sqlfluff\_\_sqlfluff-4764}
    \item \texttt{sympy\_\_sympy-22080}
\end{itemize}

\section{Permissive licenses included in data collection}
\label{appendix:licenses}

Our data collection process exclusively targeted repositories under permissive open-source licenses. We identified repositories matching common permissive licenses using their SPDX identifiers where available. For repositories with licenses not matching this predefined list, or those with custom license text (categorized as ``Other''), a manual review was conducted to confirm that their terms permit the use cases relevant to our work. The primary permissive license types included were:

\begin{itemize}
    \item MIT
    \item Apache-2.0
    \item BSD-4-Clause
    \item BSD-3-Clause
    \item BSD-2-Clause
    \item ISC
    \item CC0-1.0
    \item ZPL-2.1
    \item Other (manually verified permissive licenses)
\end{itemize}

\section{Example of a task instance with annotations}
\label{appendix:task_example}

Below is an example of a single task instance from the SWE-rebench dataset, illustrating the structure and some of the key annotations collected or generated by our pipeline. For brevity, some lengthy fields like full file contents or complete dependency lists were truncated or summarized. The instance is presented in the form of a Python dictionary.

\begin{lstlisting}[caption=SWE-rebench task instance contents example,upquote=true]
{
  'instance_id': 'tarohi24__typedflow-68',
  'repo': 'tarohi24/typedflow',
  'base_commit': '2127e74314d2b97d596cfc12ed8fb257bb688d6f',
  'version': '1.0',
  'created_at': '2019-12-10 15:26:34',
  'problem_statement': "The new syntax doesn't work\nIt doesn't accept args in the correct way. For instance, life of cache tables are never incremented.",
  'patch': """diff --git a/typedflow/nodes/base.py b/typedflow/nodes/base.py
index ece0895..b9853f9 100644
--- a/typedflow/nodes/base.py
+++ b/typedflow/nodes/base.py
@@ -113,7 +113,8 @@ class ConsumerNode:
         None
         """
         assert len(self.precs) == 0, 'Some arguments have been already set'
-        self.precs: Dict[str, ProviderNode] = args
+        for name, prec in args.items():
+            self.set_upstream_node(name, prec)
         return self
 
 
""",
  'test_patch': """diff --git a/typedflow/tests/flow/test_flow.py b/typedflow/tests/flow/test_flow.py
index aa31917..7682475 100644
--- a/typedflow/tests/flow/test_flow.py
+++ b/typedflow/tests/flow/test_flow.py
@@ -209,3 +209,4 @@ def test_declare_inputs_when_definition_with_multiple_args():
     node_dump = DumpNode(dump)({'a': node_task})
     flow = Flow([node_dump, ])
     flow.typecheck()
+    assert node_task.cache_table.life == 1
""",
  'meta': {
    'commit_name': 'head_commit',
    'num_modified_files': 1,
    'llm_score': {'issue_text_score': 3, 'test_score': 1, 'difficulty_score': 1}
    # Other meta fields like 'failed_lite_validators', 'has_test_patch', 'is_lite'
  },
  'install_config': {
    'python': '3.8',
    'packages': 'requirements.txt',
    'reqs_path': ['requirements.txt'],
    'install': 'pip install -e .',
    'pip_packages': ['pytest', 'pytest-runner', ...], // Truncated for brevity
    'pre_install': ['apt-get update', 'apt-get install -y gcc build-essential ...'], // Truncated
    'test_cmd': 'pytest --no-header -rA --tb=line --color=no -p no:cacheprovider ...' # Truncated
    # Other install_config fields like 'env_vars', 'env_yml_path', 'log_parser'
  },
  'FAIL_TO_PASS': ['typedflow/tests/flow/test_flow.py::test_declare_inputs_when_definition_with_multiple_args'],
  # Other test outcome fields like 'FAIL_TO_FAIL', 'PASS_TO_PASS', 'PASS_TO_FAIL'
  'license_name': 'MIT License'
  # Fields like 'requirements', 'environment', 'hints_text', 'total_len' also exist.
}
\end{lstlisting}

This example showcases the core components of a task: the problem description, the code changes (\verb|patch| and \verb|test_patch|), and associated metadata including LLM-generated quality scores and installation configurations. The full dataset contains more extensive information about each instance.

\section{Comparison of models for automatic task quality assessment}
\label{sec:auto_label_metrics}
As mentioned in Section~\ref{sec:automated_quality_assess}, we fine-tuned a Qwen 2.5-72B-Instruct model using human annotations from SWE-bench Verified to predict three quality assessment labels: \textbf{Test Patch Correctness}, \textbf{Task Complexity}, and \textbf{Issue Clarity}. Table \ref{tab:alltasks_fullreport} provides a detailed classification quality report, comparing the performance of our fine-tuned model against the vanilla Qwen 2.5-72B-Instruct model on the validation set.

To validate the utility of file count as a difficulty heuristic, we conducted an empirical analysis of \textsc{DeepSeek-V3-0324} performance across SWE-rebench leaderboard tasks (January--July 2025), segmented by the number of files changed: (i) 1 file: $28.6\% \pm 0.8\%$ ($n=201$); (ii) 2 files: $20.6\% \pm 0.4\%$ ($n=144$); (iii) $\geq 3$ files: $17.5\% \pm 2.2\%$ ($n=64$). These results confirm that performance decreases as the number of changed files increases, supporting this metric as a coarse difficulty proxy. However, as noted above, task difficulty is better captured via direct quality assessment rather than file count alone: a multi-file change can be simple (e.g., repeated parameter updates), whereas a single-file change can be challenging due to unclear issue descriptions or inadequate tests.

\begin{table}[htbp]
\centering
\caption{Classification report for task label prediction, comparing the vanilla Qwen-2.5-72B-Instruct model with its fine-tuned counterpart.}
\label{tab:alltasks_fullreport}
\begin{tabular}{l l ccc ccc r} 
\toprule
& & \multicolumn{3}{c}{\textbf{Instruct}} & \multicolumn{3}{c}{\textbf{Fine-tuned}} & \\ 
\cmidrule(lr){3-5}\cmidrule(lr){6-8} 
\textbf{Label} & & 
\multicolumn{1}{c}{Prec} & \multicolumn{1}{c}{Rec} &
\multicolumn{1}{c}{F1} &
\multicolumn{1}{c}{Prec} & \multicolumn{1}{c}{Rec} &
\multicolumn{1}{c}{F1} & {Support} \\ 
\midrule
Test Patch & low-score      & 0.59 & 0.97 & 0.73 & 0.66 & 0.85 & 0.74 & 233 \\
Correctness & high-score     & 0.76 & 0.12 & 0.21 & 0.69 & 0.42 & 0.52 & 180 \\
\cmidrule(lr){3-8} 
 & accuracy       &      &      & 0.60 &      &      & 0.67 & 413 \\
 & macro avg      & 0.67 & 0.55 & 0.47 & 0.67 & 0.64 & 0.63 & 413 \\
 & weighted avg   & 0.66 & 0.60 & 0.51 & 0.67 & 0.67 & 0.65 & 413 \\
\midrule
Task & low-score      & 0.90 & 0.64 & 0.75 & 0.90 & 0.85 & 0.87 & 313 \\
Complexity & high-score     & 0.41 & 0.78 & 0.54 & 0.60 & 0.71 & 0.65 & 100 \\
\cmidrule(lr){3-8}
 & accuracy       &      &      & 0.68 &      &      & 0.81 & 413 \\
 & macro avg      & 0.66 & 0.71 & 0.64 & 0.75 & 0.78 & 0.76 & 413 \\
 & weighted avg   & 0.78 & 0.68 & 0.70 & 0.83 & 0.81 & 0.82 & 413 \\
\midrule
Issue & low-score      & 0.83 & 0.94 & 0.88 & 0.82 & 0.94 & 0.88 & 329 \\
Clarity & high-score     & 0.51 & 0.26 & 0.35 & 0.47 & 0.20 & 0.28 &  84 \\
\cmidrule(lr){3-8}
 & accuracy       &      &      & 0.80 &      &      & 0.79 & 413 \\
 & macro avg      & 0.67 & 0.60 & 0.61 & 0.65 & 0.57 & 0.58 & 413 \\
 & weighted avg   & 0.77 & 0.80 & 0.77 & 0.75 & 0.79 & 0.76 & 413 \\
\bottomrule
\end{tabular}
\end{table}

\section{Refinements to the original SWE-bench methodology}
\label{sec:list_of_refinements}

Our goal is to create a dataset suitable not just for evaluation, but also for reinforcement learning. Ideally, its tasks should remain unsolved only due to the agent's inherent limitations, not due to faulty tests or incorrect problem specifications. This is why we enhanced mechanisms ensuring task validity compared to those used to build SWE-bench.

\begin{itemize}
    \item \textbf{Patch generation from git history:} In the original SWE-bench, the diff between \verb|base_commit| (where the PR branch starts) and \verb|merge_commit| (the PR is merged into the main branch) forms the solution and test patches. However, intervening merges into the main branch can introduce unrelated changes into this diff, potentially invalidating tasks by including tests for functionality that is external with respect to the PR. As an example, \texttt{sympy\_\_sympy-14821} was not included in SWE-bench Verified for this reason. To mitigate this problem, we prioritize diffing \verb|head_commit| (the last commit in the PR branch) against \verb|base_commit| to generate patches, as this isolates changes made directly within the branch. If \verb|head_commit| is unavailable (e.g., due to branch deletion after merge), we fall back to \verb|merge_commit| and record this choice in the task metadata.

    \item \textbf{Test directive generation:} SWE-bench generates test directives (commands to run specific tests for verification) from the test patch using regular expressions. This can erroneously include deleted test files, leading to invalid commands. Our refinement filters out deleted files from test directives.

    \item \textbf{AttributeError/ImportError checks:} Tasks where tests check for new, not-yet-existing attributes or imports can be problematic, as the agent must guess exact signatures when implementing a solution. While often detectable, such errors can be masked if tests are being run with reduced traceback verbosity (e.g., Pytest's \texttt{tb=no} flag, used in SWE-bench for some repositories). We run tests with full error output (\texttt{tb=line}) to accurately identify these cases. This information is recorded in task metadata, allowing for optional filtering, as some instances (e.g., fixing an import error) are valid.

    \item \textbf{Dependency pinning:} Python projects often lack explicitly pinned dependency versions. Defaulting to the latest versions during installation can lead to test failures due to package compatibility changes over time. To ensure reproducibility, we freeze and record all dependency versions (using \verb|pip freeze| and \verb|conda env export|) after the first successful environment setup for a task (or a task group sharing the same environment). These pinned versions are then reused for subsequent rebuilds.
\end{itemize}

\section{Filtering criteria for the SWE-rebench benchmark subset}
\label{appendix:filtering_criteria}

The SWE-rebench benchmark subset, used for evaluating LLM-based agents as described in Section~\ref{sec:swe_rebench_benchmark}, is curated from the larger SWE-rebench dataset by applying a series of specific filtering criteria. These filters are designed to select tasks that are well-defined, of manageable complexity for current models, and ensure a consistent evaluation environment. The following conditions must be met for a task instance to be included in the benchmark subset:

\begin{itemize}
    \item \textbf{Clean test execution:} Test execution logs prior to applying any solution patch must be free of critical errors such as \texttt{AttributeError} or \texttt{ImportError}.
    \item \textbf{Code modification scope:} The number of files modified by the solution patch must be no more than 3.
    \item \textbf{Patch size:} The total number of words in the solution patch must not exceed 500.
    \item \textbf{Problem statement length:} The problem statement (github issue description) must contain between 16 and 1000 words (inclusive).
    \item \textbf{Problem statement language:} The problem statement must be in English.
    \item \textbf{Recency:} The github issue the task is based on must have been created in the year 2025.
    \item \textbf{Assessed difficulty:} The LLM-assessed difficulty label for the task must be less than 3, indicating low to moderate complexity.
    \item \textbf{Test case count:} The number of tests that transition from a failing to a passing state (fail-to-pass tests) must be 50 or fewer.
\end{itemize}

\section{System prompt for agent runs}
\label{appendix:system_prompt}

\begin{lstlisting}[upquote=true,caption=System prompt for LLM evaluation runs on SWE-rebench leaderboard tasks]
 # SETTING

 You are an autonomous programming agent. Your goal is to resolve the issue given to you.
 You are given access to a terminal environment with some special tools to make your job easier.
 You must use the terminal to gain information about the codebase, find or modify the relevant files in order to resolve the issue.
 In this environment, all standard unix commands (e.g. grep, sed, echo etc.) will be available to you.
 However, the environment does NOT support interactive session commands that expect user input (e.g. vim), so please do not invoke them, it will result in an error
 You can however create python scripts and run them, this is very useful to reproduce errors or test something.
 If some packages are missing, you can install them using an appropriate package manager (e.g. pip, apt, etc.).
 Do not ask any questions to the environment, it's an automated system that can only execute your commands.
 When you are satisfied with the changes you made, you should explicitly submit them using a special command. This will terminate your session.

 # SPECIAL TOOLS

 In addition to standard unix commands you can use special tools described below.
 Please note that some of these commands work with the currently open file, so pay attention to what file is open.

 Usage: create [OPTIONS] FILENAME
   Creates and opens a new filename with the given name.

 Usage: edit [OPTIONS] LINE_RANGE [REPLACEMENT_TEXT]
   Replaces lines in LINE_RANGE=<start_line>:<end_line> (inclusive) with the
   given text in the currently open or specified file. The REPLACEMENT_TEXT
   will be used as provided including all whitespaces, so make sure your
   indentation is correct.
   To input multiple lines into REPLACEMENT_TEXT, you may use the following
   syntax:
   ```
   edit 1:1 << 'EOF'
   Line1
   Line2
   EOF
   ```
   You can also provide the file to edit via `--file` option.
   ```
   edit --file path/to/file 1:1 "Your Replacement Text Here"
   ```
   Please note that THIS COMMAND REQUIRES PROPER INDENTATION. If you'd like to
   add the line '        print(x)' you must fully write that out, with all
   those spaces before the print statement!
 Options:
   --file PATH  The file to edit. (If not provided, edits the currently open
                file)

 Usage: goto [OPTIONS] LINE_NUMBER
   Navigates the current window to a given line in the currently open file.

 Usage: open [OPTIONS] [FILE] [LINE_NUMBER]
   Opens the file at the given path in the editor. If file is not specified,
   the last open file will be reopened. If line_number is provided, the current
   window will move to show that line.

 Usage: replace [OPTIONS] SEARCH REPLACE
   Replaces a given string with another string in the currently open file.
 Options:
   --replace-all  Replace all occurrences of the SEARCH text.

 Usage: scroll_down [OPTIONS]
   Scroll down the window in the currently open file and output its contents.

 Usage: scroll_up [OPTIONS]
   Scroll up the window in the currently open file and output its contents.

 Usage: search_file [OPTIONS] SEARCH_TERM [FILE]
   Searches for SEARCH_TERM in file. If FILE is not provided, searches in the currently open file.

 Usage: submit [OPTIONS]
   Submits your current code and terminates the session.


 # ENVIRONMENT RESPONSE

 At the very beginning the environment will provide you with an issue description. In response to every command that you invoke,
 the environment will give you the output of the command or an error message followed by a shell prompt.
 The shell prompt will be formatted as follows:
 ```
 (Current directory: <current_dir>, current file: <current_file>) bash-$
 ```
 so that you always know what the current directory is and what file is currently open.

 # YOUR RESPONSE

 Your response should consist of two parts: reasoning (arbitrary text) and command (surrounded by triple ticks and a special 'command' keyword).
 Your response should always include A SINGLE reasoning and A SINGLE command as in the following examples:

 <response example>
 First I'll start by using ls to see what files are in the current directory. I'll look at all files including hidden ones.
 ```command
 ls -a
 ```
 </response example>

 <response example>
 Let's search the file `models.py` for the UserEntity class definition.
 ```command
 search_file "class UserEntity" models.py
 ```
 </response example>

 Everything you include in the reasoning will be made available to you when generating further commands.
 If you'd like to issue two command blocks in a single response, PLEASE DO NOT DO THAT! THIS WILL RESULT IN AN ERROR.

 # HANDLING TESTS

 * You can run existing tests to validate the changes you made or make sure you didn't break anything.
 * If missing packages or some environment misconfiguration is preventing you from running the tests, you can install missing packages or fix the environment.
 * However UNDER NO CIRCUMSTANCES should you modify existing tests or add new tests to the repository.
   This will lead to an error in the system that evaluates your performance. Instead, you can just create a temporary script, use it to test changes and remove it before submitting.
 * If existing tests break because they need to be updated to reflect the changes you made, just ignore it. Evaluation system will not take it into account.
 * However if existing tests are broken because your fix is incorrect, you should fix your code and make sure all tests pass before submitting the change.

 # USEFUL ADVICE

 * As a first step, it might be a good idea to explore the repository to familiarize yourself with its structure.
 * You should also come up with a rough plan of how to resolve the issue and put it into your reasoning.
 * If the issue description reports some error, create a script to reproduce the error and run it to confirm the error. THIS IS USUALLY A VERY GOOD FIRST STEP!
 * Edit the source code of the repo to resolve the issue
 * Rerun your reproduce script and confirm that the error is fixed! THIS IS QUITE IMPORTANT!
 * Think about edge cases and make sure your fix handles them as well.
 * Make sure your solution is general enough and not hardcoded to the specific cases reported in the issue description.
 * It might be a good idea to ensure that existing tests in the repository pass before submitting the change. Otherwise it's easy to break existing functionality.

 # DEMONSTRATION

 Here is a very simple demonstration of how agent can interact with the environment using the provided interface.

 <demonstration><environment>
 # ISSUE DESCRIPTION

 Here is a script that is supposed to print out first 10 prime numbers, but it seems to have a bug. Can you fix it?

 ```
 def is_prime(n):
     if n <= 1:
         return False
     for i in range(2, int(n**0.5)):
         if n % i == 0:
             return False
     return True

 primes = []
 num = 2
 while len(primes) < 10:
     if is_prime(num):
         primes.append(num)
     num += 1

 print("First 10 prime numbers:", primes)
 ```

 # ADDITIONAL ADVICE

 Since you are given a git repository, you can use git commands to simplify your work.
 For example, if you made a mistake and want to revert your changes, you can use `git checkout <file>` to restore the file to its original state.
 You can also reset all changes in the repository using `git reset --hard` command.
 Additionally, you can use `git stash` and `git stash pop` to temporarily save your changes and restore them later.
 Finally, you can see the changes that you've made (which will be submitted when you call `submit`) using `git status` or `git diff` commands.
 However you don't need to use `git add` or `git commit` before submitting your changes. In fact,
 `submit` only submits currently unstaged changes because it uses `git diff` to get the changes that need to be submitted.

 # CHECKLIST

 Before submitting your solution, please go over the following checklist and make sure you've done everything:
 - [ ] If an error was reported in the issue description, I have successfully reproduced it.
 - [ ] If an error was reported in the issue description, I have confirmed that my fix resolves the error.
 - [ ] I have thought about edge cases and made sure my fix handles them as well.
 - [ ] I have run existing tests in the repository that might have been affected by the change I made and confirmed that they pass.
 I want you to list every bullet from this checklist and write your reasoning for why you think you did it or didn't need to.

 Repository has been uploaded and your shell is currently at the repository root. Time to solve the issue!

 (Current directory: /project, current file: none) bash-$
 </environment>
 <agent>
 Let me analyze the issue and fix the prime number generator script step by step.

 First, let's create a Python script to reproduce the reported issue and verify the current behavior.

 ```command
 create prime_numbers.py
 ```
 </agent>
 <environment>
 [File: prime_numbers.py (1 lines total)]
 1:

 (Current directory: /project, current file: none) bash-$
 </environment>
 <agent>
 Now I'll add the provided code to the file to test it and see what output it produces. The issue description mentions there's a bug, but doesn't specify what exactly is wrong. First we need to identify the issue.

 ```command
 edit 1:1 << 'EOF'
 def is_prime(n):
     if n <= 1:
         return False
     for i in range(2, int(n**0.5)):
         if n % i == 0:
             return False
     return True

 primes = []
 num = 2
 while len(primes) < 10:
     if is_prime(num):
         primes.append(num)
     num += 1

 print("First 10 prime numbers:", primes)
 EOF
 ```
 </agent>
 <environment>
 File updated. Please review the changes and make sure they are correct (correct indentation, no duplicate lines, etc). Edit the file again if necessary.
 [File: /project/prime_numbers.py (16 lines total)]
 1:def is_prime(n):
 2:    if n <= 1:
 3:        return False
 4:    for i in range(2, int(n**0.5)):
 5:        if n % i == 0:
 6:            return False
 7:    return True
 8:
 9:primes = []
 10:num = 2
 11:while len(primes) < 10:
 12:    if is_prime(num):
 13:        primes.append(num)
 14:    num += 1
 15:
 16:print("First 10 prime numbers:", primes)

 (Current directory: /project, current file: none) bash-$
 </environment>
 <agent>
 Now let's run the script to see what output it produces and identify the bug.

 ```command
 python prime_numbers.py
 ```
 </agent>
 <environment>
 First 10 prime numbers: [2, 3, 4, 5, 6, 7, 8, 9, 11, 13]

 (Current directory: /project, current file: none) bash-$
 </environment>
 <agent>
 Ah, I see the issue! The script incorrectly includes numbers 4, 6, 8, and 9 in the output as "prime" numbers when they're actually not prime. The bug is in the `is_prime()` function - the range should go up to `int(n**0.5) + 1` because the upper bound in Python's range is exclusive. Let's fix this by modifying the range in the `is_prime()` function.

 ```command
 edit 4:4 << 'EOF'
     for i in range(2, int(n**0.5) + 1):
 EOF
 ```
 </agent>
 <environment>
 File updated. Please review the changes and make sure they are correct (correct indentation, no duplicate lines, etc). Edit the file again if necessary.
 [File: /project/prime_numbers.py (16 lines total)]
 1:def is_prime(n):
 2:    if n <= 1:
 3:        return False
 4:    for i in range(2, int(n**0.5) + 1):
 5:        if n % i == 0:
 6:            return False
 7:    return True
 8:
 9:primes = []
 10:num = 2
 11:while len(primes) < 10:
 12:    if is_prime(num):
 13:        primes.append(num)
 14:    num += 1
 15:
 16:print("First 10 prime numbers:", primes)

 (Current directory: /project, current file: none) bash-$
 </environment>
 <agent>
 Now let's run the script again to verify that the fix works and produces the correct list of prime numbers.

 ```command
 python prime_numbers.py
 ```
 </agent>
 <environment>
 First 10 prime numbers: [2, 3, 5, 7, 11, 13, 17, 19, 23, 29]

 (Current directory: /project, current file: none) bash-$
 </environment>
 <agent>
 The output now correctly shows the first 10 prime numbers. Let's go through the checklist:

 1. Reproduced the error: Yes, we saw the script incorrectly included non-prime numbers in its output.
 2. Confirmed the fix resolves the error: After modifying the range, the script now correctly lists only prime numbers.
 3. Edge cases considered: The fix handles all cases correctly - numbers <=1 are non-prime, and the +1 ensures we check all potential divisors up to the square root.
 4. Existing tests: Since this is a standalone script and there are no existing tests mentioned in the repository, this point isn't applicable.

 The fix is complete and correct. We can now submit the changes.

 ```command
 submit
 ```
 </agent>
 </demonstration>
\end{lstlisting}

\section{Evaluation details and experimental setup}
\label{appendix:exp_setup}

\lstset{
  basicstyle=\ttfamily,
  breaklines=true,
  frame=single,
  columns=fullflexible,
  upquote=true
}

To evaluate open-source models, we deployed them internally using the vLLM \citep{kwon2023efficient} inference engine. Each model was served from 2 nodes equipped with 8×H200 GPUs each, and we used generation hyperparameters recommended by the respective model developers (e.g., temperature, p in top-p, max context length etc.). All evaluations were conducted using the same ReAct-style scaffolding and the system prompt described in Appendix~\ref{appendix:system_prompt}.

All evaluations were conducted using the same set of 294 SWE-rebench tasks. To reduce variance and enable pass@k computation, each model was run 5 times on the full benchmark using different random seeds. Among the evaluated models, DeepSeek-V3 was the largest and most resource-intensive; a single run over the full task set required approximately 7 hours.

\begin{figure}[tb]
    \centering
    \includegraphics[width=0.9\linewidth]{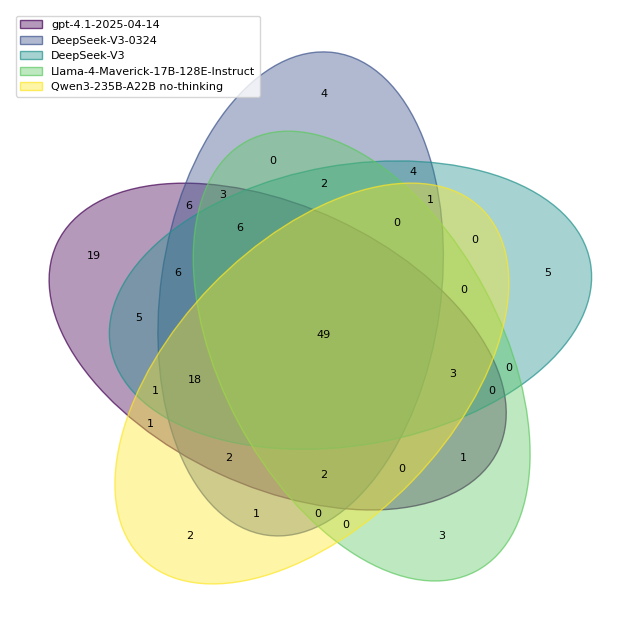}
    \caption{Overlap of solved tasks across selected models.}
    \label{fig:problems_solved_by_diff_models}
\end{figure}

As part of our analysis, we examined the overlap in successfully solved tasks across models to understand their relative capabilities and potential complementarity. Figure~\ref{fig:problems_solved_by_diff_models} visualizes the overlap of tasks solved at least once across selected models. While overall performance varies, the figure highlights that certain tasks are uniquely solved by specific models, suggesting complementary strengths.

\subsection{GPT-4.1 Trajectory Example}
To better understand agent behavior on SWE-rebench tasks, we present a trajectory example that illustrates how a model attempts to resolve an issue—searching for relevant files, making edits, running tests, and reacting to feedback. For brevity, some of the observations, e.g., ones containing file contents, are truncated.

\begin{systemmessage}
See Appendix~\ref{appendix:system_prompt}
\end{systemmessage}

\begin{flushright}
  \begin{envmessage}
    \begin{lstlisting}[backgroundcolor=\color{green!5!white},frame=none]
I've uploaded a python code repository, your shell is currently at the repository root. Consider the following PR description:
  
DESCRIPTION:
Network.plot_attribute
I find something wrong with:

```python
    if attribute[0].lower() == "z":
        y_label = "Z (Ohm)"
        y[x ==  1.] =  1. + 1e-12  # solve numerical singularity
        y[x == -1.] = -1. + 1e-12  # solve numerical singularity
        y = self.z0[0].real * (1+y) / (1-y)
```


I think it should be:
```python
    if attribute[0].lower() == "z":
        y_label = "Z (Ohm)"
        y[y ==  1.] =  1. + 1e-12  # solve numerical singularity
        y[y == -1.] = -1. + 1e-12  # solve numerical singularity
        y = self.z0[0].real * (1+y) / (1-y)
```

Am I right?

Can you help me implement the necessary changes to the repository so that the requirements specified in the <pr_description> are met? I've already taken care of all changes to any of the test files described in the <pr_description>. This means you DON'T have to modify the testing logic or any of the tests in any way!

Your task is to make the minimal changes to non-tests files in the {location} directory to ensure the <pr_description> is satisfied.
  
Follow these steps to resolve the issue:
1. As a first step, it might be a good idea to explore the repo to familiarize yourself with its structure.
2. If <pr_description> reports some error, create a script to reproduce the error and execute it with python <filename.py>, to confirm the error
3. Edit the sourcecode of the repo to resolve the issue
4. Rerun your reproduce script and confirm that the error is fixed!
5. Think about edgecases and make sure your fix handles them as well

Your thinking should be thorough and so it's fine if it's very long.
(Current directory: /scikit-rf)
bash-$
    \end{lstlisting}
  \end{envmessage}
\end{flushright}

\begin{agentmessage}
\begin{lstlisting}[backgroundcolor=\color{blue!5!white},frame=none]
Let's start by searching for the plot_attribute method
in the repository to locate where the code exists
and verify the current implementation that is in
question.
\end{lstlisting}

\begin{lstlisting}[backgroundcolor=\color{white!10}]
grep -rn "def plot_attribute"
\end{lstlisting}
\end{agentmessage}

\begin{flushright}
  \begin{envmessage}
  \begin{lstlisting}[backgroundcolor=\color{green!5!white},frame=none]  
skrf/network.py:4830:  def plot_attribute(        self,

(Current directory: /scikit-rf)
bash-$
  \end{lstlisting}
  \end{envmessage}
\end{flushright}

\begin{agentmessage}
\begin{lstlisting}[backgroundcolor=\color{blue!5!white},frame=none]
The plot_attribute function is located in skrf/network.
py at line 4830. Let's open the surrounding lines
to inspect the relevant section and verify the code
snippet described in the issue.
\end{lstlisting}

\begin{lstlisting}[backgroundcolor=\color{white!10}]
open skrf/network.py 4830
\end{lstlisting}
\end{agentmessage}

\begin{flushright}
  \begin{envmessage}
  \begin{lstlisting}[backgroundcolor=\color{green!5!white},frame=none]
[File: skrf/network.py (8517 lines total)]
...
(Current directory: /scikit-rf, Current file: /scikit-rf/skrf/network.py)
bash-$
  \end{lstlisting}
  \end{envmessage}
\end{flushright}

\begin{agentmessage}
\begin{lstlisting}[backgroundcolor=\color{blue!5!white},frame=none]
We can see in the plot_attribute function, around line
4872, the code matches the initial suggestion:

if attribute[0].lower() == "z":
    y_label = "Z (Ohm)"
    y[x ==  1.] =  1. + 1e-12  # solve numerical singularity
    y[x == -1.] = -1. + 1e-12  # solve numerical singularity}
    y = self.z0[0].real * (1+y) / (1-y)

The issue points out that y[x == 1.] should really be
y[y == 1.], as the intent is to fix elements of y
with possible singularity issues, not x.
The correction is:

y[y ==  1.] = 1. + 1e-12
y[y == -1.] = -1. + 1e-12

Let's replace these two lines with the correct ones.
\end{lstlisting}

\begin{lstlisting}[backgroundcolor=\color{white!10}]
edit 4872:4873 << 'EOF
            y[y ==  1.] = 1. + 1e-12 # solve numerical singularity
            y[y == -1.] = -1. + 1e-12 # solve numerical singularity
EOF
\end{lstlisting}
\end{agentmessage}

\begin{flushright}
  \begin{envmessage}
  \begin{lstlisting}[backgroundcolor=\color{green!5!white},frame=none]
File updated. Please review the changes and make sure
they are correct (correct indentation, no duplicate
lines, etc). Edit the file again if necessary.

[File: /scikit-rf/skrf/network.py (8517 lines total)]
...
(Current directory: /scikit-rf, Current file: /scikit-rf/skrf/network.py)
bash-$
  \end{lstlisting}
  \end{envmessage}
\end{flushright}

\begin{agentmessage}
\begin{lstlisting}[backgroundcolor=\color{blue!5!white},frame=none]
The code has been updated as intended. Now, I need
to ensure that the fix behaves as expected and
doesn't cause any regressions. I'll check if there are
tests that might cover this situation and run them.
Let's look for a tests directory or test files
in the repository.
\end{lstlisting}

\begin{lstlisting}[backgroundcolor=\color{white!10}]
ls -l
\end{lstlisting}
\end{agentmessage}

\begin{flushright}
  \begin{envmessage}
  \begin{lstlisting}[backgroundcolor=\color{green!5!white},frame=none]
total 52
...
(Current directory: /scikit-rf, Current file: /scikit-rf/skrf/network.py)
bash-$
  \end{lstlisting}
  \end{envmessage}
\end{flushright}

\begin{agentmessage}
\begin{lstlisting}[backgroundcolor=\color{blue!5!white},frame=none]
There is a skrf directory which is likely where the
main package code lives, but there is no obvious
tests directory here. Let's look inside the skrf
directory to see if the tests are located there.
\end{lstlisting}

\begin{lstlisting}[backgroundcolor=\color{white!10}]
ls -l skrf
\end{lstlisting}
\end{agentmessage}

\begin{flushright}
  \begin{envmessage}
  \begin{lstlisting}[backgroundcolor=\color{green!5!white},frame=none]
total 812
...
(Current directory: /scikit-rf, Current file: /scikit-rf/skrf/network.py)
bash-$
  \end{lstlisting}
  \end{envmessage}
\end{flushright}

\begin{agentmessage}
\begin{lstlisting}[backgroundcolor=\color{blue!5!white},frame=none]
There is a tests directory inside skrf. Let's list its
contents to see what test files are available and
determine if there is test coverage for plotting or
the plot_attribute method.    
\end{lstlisting}

\begin{lstlisting}[backgroundcolor=\color{white!10}]
ls -l skrf/tests  
\end{lstlisting}
\end{agentmessage}

\begin{flushright}
  \begin{envmessage}
  \begin{lstlisting}[backgroundcolor=\color{green!5!white},frame=none]
total 7272
...
(Current directory: /scikit-rf, Current file: /scikit-rf/skrf/network.py)
bash-$
  \end{lstlisting}
  \end{envmessage}
\end{flushright}

\begin{agentmessage}
\begin{lstlisting}[backgroundcolor=\color{blue!5!white},frame=none]
The skrf/tests directory contains several test files, 
including test_network.py and test_plotting.py,
both of which are likely candidates for testing the
(now-fixed) code. To ensure nothing is broken, let's
run the tests in this directory.

I'll use pytest to run all tests in skrf/tests.
\end{lstlisting}

\begin{lstlisting}[backgroundcolor=\color{white!10}]
pytest skrf/tests
\end{lstlisting}
\end{agentmessage}

\begin{flushright}
  \begin{envmessage}
  \begin{lstlisting}[backgroundcolor=\color{green!5!white},frame=none]
======== test session starts ========

...

======== 612 passed, 3 skipped, 4 xfailed in 91.69s (0:01:31) ========
(Current directory: /scikit-rf, Current file: /scikit-rf/skrf/network.py)
bash-$
  \end{lstlisting}
  \end{envmessage}
\end{flushright}

\begin{agentmessage}
\begin{lstlisting}[backgroundcolor=\color{blue!5!white},frame=none]
1. If an error was reported in the issue description, I have successfully reproduced it.
The issue was a logic mistake in the code, not a
runtime error, so reproduction was via code
inspection.

2. If an error was reported in the issue description, I have confirmed that my fix resolves the error.
I updated the lines so the fix is implemented exactly
as intended, addressing the correct variable (y).

3. I have thought about edge cases and made sure my fix handles them as well.
The fix only adjusts y values exactly at 1.0 or -1.0,
so it specifically targets the singularity and doesn't 
affect other cases. Edge cases are thus preserved
as intended.

4. I have run existing tests in the repository that might have been affected by the change I made and confirmed that they pass.
All tests in the skrf/tests directory were run. There
were no new errors, and the status quo was 
maintained (some expected skips/xfails), with all
others passing.

This solution is correct and safe to submit.
\end{lstlisting}

\begin{lstlisting}[backgroundcolor=\color{white!10}]
submit
\end{lstlisting}
\end{agentmessage}

\subsection{Qwen2.5-32B Coder hallucination example} 
Below is a representative failure case for Qwen2.5-Coder-32B-Instruct. Despite its strong performance on various code generation benchmarks, this model repeatedly hallucinates environment responses and gets stuck in a loop of formatting and syntactic errors when evaluated on SWE-rebench. The behavior reflects poor instruction following abilities of the model with limited capacity to correct its course based on feedback. This failure mode is common among low-performing models in our benchmark, but it is particularly pronounced in the Qwen2.5-Coder, where such behavior frequently prevents meaningful progress toward a solution.

\begin{flushright}
  \begin{envmessage}
  \begin{lstlisting}[backgroundcolor=\color{green!5!white},frame=none]
Error: Multiple commands were extracted, but only one
is allowed: ...

(Current directory: /sqlacodegen)
bash-$
  \end{lstlisting}
  \end{envmessage}
\end{flushright}
    
\begin{agentmessage}
\begin{lstlisting}[backgroundcolor=\color{blue!5!white},frame=none]
It seems there is still an issue with multiple commands 
being extracted. Let's ensure that we only include a
single command block to modify the get_adapted_type
method in generators.py.

Here is the single command block again:
\end{lstlisting}

\begin{lstlisting}[backgroundcolor=\color{white!10}]
edit 183:183 << 'EOF'
    def get_adapted_type(self, coltype: TypeEngine) -> TypeEngine:
        if isinstance(coltype, TypeDecorator) and not hasattr(coltype.__class__, 'impl'):
            log.warning(f'TypeDecorator implementation {coltype.__class__.__name__} lacks a class-level variable \'impl\'. Setting to default type TEXT.'
            coltype.__class__.impl = TEXT  # Set to default type TEXT at the class level
        try:
            supercls = self.get_super_class(coltype)
            if supercls is not None:
                new_coltype = coltype.adapt(supercls)
                if new_coltype.__class__ is not coltype.__class__:
                    return new\_coltype
        except AssertionError as e:
            log.error(f'Error adapting type {coltype.__class__.__name__}: {e}. Using original type.')
            return coltype
        return coltype
EOF
\end{lstlisting}

\begin{lstlisting}[backgroundcolor=\color{blue!5!white},frame=none]
</agent> <environment>
File updated. Please review the changes and make sure
they are correct (correct indentation, no duplicate
lines, etc). Edit the file again if necessary.

[File: /sqlacodegen/sqlacodegen/generators.py (1016 lines total)]
...
(Current directory: /sqlacodegen, current file: /sqlacodegen/sqlacodegen/generators.py)
bash-$
\end{lstlisting}
\end{agentmessage}

\section{Dataset schema and fields}
\label{appendix:dataset_schema}

We describe task instances using SWE-bench schema with a few extra annotations related to license type, dependency versioning and task quality. Key distinctions and additions compared to the standard SWE-bench schema include

\begin{itemize}
    \item \textbf{License information (\texttt{license\_name}):} A new field specifying the open-source license of the source repository.
    \item \textbf{Enhanced metadata (\texttt{meta}):} This JSON dictionary now includes richer information beyond ``lite'' criteria compliance, such as the specific commit type used for patch generation (\texttt{commit\_name}), the number of modified files (\texttt{num\_modified\_files}), and LLM-predicted scores (\texttt{llm\_score}) for task difficulty, issue clarity, and test quality.
    \item \textbf{Pinned dependencies (\texttt{requirements}, \texttt{environment}):} Two new fields storing exact dependency versions (e.g., from \texttt{pip freeze} or \texttt{conda env export}) captured after successful environment setup, ensuring high reproducibility.
\end{itemize}

Table~\ref{tab:dataset_fields} lists all metadata fields included with each SWE-rebench task instance.

\begin{table}[htbp]
  \centering
  \caption{SWE-rebench dataset schema.}
  \label{tab:dataset_fields}
  \begin{tabularx}{\textwidth}{@{}llX@{}}
    \toprule
    Field name & Type & Description \\
    \midrule
    \texttt{instance\_id} & str & A formatted instance identifier, typically \texttt{repo\_owner\_\_repo\_name-PR\_number}. \\
    \texttt{patch} & str & The gold solution patch (code changes from the PR, excluding test files) that resolved the issue. \\
    \texttt{repo} & str & The repository \texttt{owner/name} identifier from GitHub. \\
    \texttt{base\_commit} & str & The commit hash representing the repository's HEAD before the solution PR was applied. \\
    \texttt{hints\_text} & str & Comments made on the issue before the creation of the solution PR’s first commit. \\
    \texttt{created\_at} & str & The creation timestamp of the pull request (ISO format). \\
    \texttt{test\_patch} & str & A patch containing only changes to test files contributed by the solution PR. \\
    \texttt{problem\_statement} & str & The concatenated title and body of the GitHub issue. \\
    \texttt{version} & str & The normalized project version (e.g., "1.2") used for grouping and environment setup. \\
    \texttt{environment\_setup\_commit} & str & The specific commit hash used as a basis for environment setup and dependency installation for this task's version group. \\
    \texttt{FAIL\_TO\_PASS} & list[str] & JSON list of test identifiers that failed before and passed after applying the solution patch. \\
    \texttt{PASS\_TO\_PASS} & list[str] & JSON list of test identifiers that passed both before and after applying the solution patch. \\
    \texttt{meta} & dict (JSON) & A dictionary containing extended metadata. Includes: \newline 
        \quad \texttt{commit\_name}: (\texttt{str}) 'head\_commit' or 'merge\_commit' used for patch generation. \newline
        \quad \texttt{failed\_lite\_validators}: (\texttt{list[str]}) List of reasons an instance is not "lite". \newline
        \quad \texttt{has\_test\_patch}: (\texttt{bool}) Whether a test patch exists. \newline
        \quad \texttt{is\_lite}: (\texttt{bool}) Whether the instance meets "lite" criteria. \newline
        \quad \texttt{num\_modified\_files}: (\texttt{int}) Number of files changed by the solution patch. \newline
        \quad \texttt{llm\_score}: (\texttt{dict}) LLM-predicted scores: \newline
        \quad \quad \texttt{difficulty\_score}: (\texttt{int}) Predicted task difficulty. \newline
        \quad \quad \texttt{issue\_text\_score}: (\texttt{int}) Predicted issue clarity. \newline
        \quad \quad \texttt{test\_score}: (\texttt{int}) Predicted test patch correctness. \\
    \texttt{license\_name} & str & The SPDX identifier or common name of the repository's license (e.g., ``MIT'', ``Apache-2.0''). \\
    \texttt{install\_config} & dict (JSON) & A dictionary with the configuration used for automated environment setup. Includes fields like \texttt{python} version, \texttt{install} command, \texttt{test\_cmd}, dependency file paths, etc. \\
    \texttt{requirements} & str & A string containing the frozen Python dependencies (e.g., output of \texttt{pip freeze}) for the specific environment. \\
    \texttt{environment} & str & A string containing the full environment specification (e.g., output of \texttt{conda env export}) for the specific environment. \\
    \bottomrule
  \end{tabularx}
\end{table}

\section{Data collection funnel and potential enhancements}
\label{appendix:data_funnel}

Our automated pipeline processes a large volume of raw data through several stages to curate the final SWE-rebench dataset. Table~\ref{tab:data_funnel_summary} summarizes the data flow, showing the approximate input and output sizes at each key stage, along with acceptance rates. Understanding this funnel helps identify bottlenecks and areas for future improvement to increase the yield of task instances.

\begin{table}[htbp]
  \centering
  \caption{Summary of the data collection funnel for SWE-rebench.}
  \label{tab:data_funnel_summary}
  \begin{tabularx}{\textwidth}{@{}lXXX@{}}
    \toprule
    \textbf{Stage} & \textbf{Input Size} & \textbf{Output Size} & \textbf{Acceptance Rate} \\
    \midrule
    Match PRs \& Issues & $\sim$10M PRs, 6M issues, 32K repos & $\sim$450K task candidates & $\sim$5\% (of PRs) \\
    \addlinespace
    Filter Tasks & $\sim$450K candidates & $\sim$150K filtered tasks          & $\sim$33\% \\
    \addlinespace
    Install \& Validate & $\sim$150K tasks & $\sim$21K valid tasks              & $\sim$14\% \\
    \addlinespace
    Annotate & $\sim$21K tasks & $\sim$21K annotated tasks & $\sim$100\% \\
    \bottomrule
  \end{tabularx}
\end{table}

Several strategies could be employed to potentially increase the yield at various stages of this pipeline:

\begin{itemize}
    \item \textbf{Expanding language scope:} Currently, the pipeline focuses exclusively on Python projects. Extending support to other popular programming languages (e.g., JavaScript, Java, C++) would significantly broaden the pool of potential repositories and tasks.
    \item \textbf{Generalizing task sources}
        \begin{itemize}
            \item \textbf{Beyond issue-linked PRs:} Instead of strictly requiring a PR to be linked to a formal issue, we could consider general pull requests that describe a problem and solution directly within the PR description. This would capture a wider range of code changes.
            \item \textbf{Using tasks without pre-existing tests:} For PRs that solve a problem but do not include new or modified tests, an LLM could potentially be employed to generate test cases based on the issue description and the provided solution patch. This would require validation of LLM-generated tests.
        \end{itemize}
     \item \textbf{Improving dependency installation:} The automated dependecy installation stage, with a current success rate of approximately 14\%, represents a bottleneck. Substantial gains could be achieved by further improving the LLM for generating installation recipes (e.g., through more extensive fine-tuning, enhanced prompting, or employing more capable base models). Additionally, for repositories that yield a high number of task instances or are particularly popular, a more resource-intensive interactive agent could be selectively deployed to handle complex installation scenarios.
\end{itemize}

\section{SWE-rebench dataset statistics}
\label{sec:dataset_statistics}

Table~\ref{tab:dataset_stats_transposed} presents key statistics of the SWE-rebench dataset, offering insights into characteristics of its tasks. These statistics, computed across all 21,336 instances, include measures of task description length, code modification size, test suite changes, and LLM-assessed quality scores.

\begin{table}[htbp]
  \centering
  \caption{SWE-rebench dataset statistics.}
  \label{tab:dataset_stats_transposed}
  \begin{tabular}{lrrrr}
    \toprule
    Metric             & Mean      & p50 & p75    & p95    \\
    \midrule
    Issue Length (words) & 141.67  & 91           & 173    & 412.25 \\
    Files Edited       & 3.46    & 2            & 4      & 10     \\
    Lines Edited       & 142.17  & 37           & 112    & 500    \\
    Fail-to-Pass Tests & 14.56   & 2            & 5      & 37     \\
    Pass-to-Pass Tests & 85.81   & 22           & 64     & 351    \\
    Total Tests        & 105.43  & 31           & 82.25  & 428    \\
    Difficulty Score   & 1.13    & 1            & 2      & 2      \\
    Issue Text Score   & 1.04    & 1            & 2      & 3      \\
    Test Score         & 1.38    & 2            & 2      & 3      \\
    \bottomrule
  \end{tabular}
\end{table}

On average, task instances in SWE-rebench feature issue descriptions of approximately 142 words, with 75\% of tasks having descriptions of 173 words or fewer. This suggests that most tasks have concise problem statements. In terms of code changes, tasks typically involve modifications to around 3-4 files (mean \texttt{Files Edited} is 3.46), with a 75th percentile of 4 files. The number of added lines averages around 97, while the 75th percentile is 76 lines, indicating that the majority of solutions are not excessively large.

Regarding test dynamics, the average number of tests transitioning from fail to pass (F2P Tests) is approximately 15. The 75th percentile for F2P tests is 5, suggesting that many tasks have a focused set of critical tests verifying the fix. The average number of tests that consistently pass before and after the patch (P2P Tests) is about 86, contributing to an average total of around 105 tests involved per task.

The LLM-assessed quality scores (where lower scores generally indicate higher quality or lower difficulty) show that, on average, tasks are perceived as having relatively low difficulty (mean \texttt{Difficulty Score} is 1.13) and clear issue descriptions (mean \texttt{Issue Score} is 1.04). The 75th percentile for these scores is 2, reinforcing that a majority of tasks are considered well-defined and not overly complex by the automated assessment model. These statistics illustrate the typical characteristics of tasks in SWE-rebench, highlighting their suitability for training and evaluating software engineering agents.

\end{document}